\documentclass[reprint, amsmath, amssymb, aps, superscriptaddress]{revtex4-1}
\usepackage{format}
\allowdisplaybreaks

\begin{document}

\preprint{APS/123-QED}

\title{ Prospects for thermalization of microwave-shielded ultracold molecules } 

\author{Reuben R. W. Wang} 
\author{John L. Bohn}
\affiliation{ JILA, NIST, and Department of Physics, University of Colorado, Boulder, Colorado 80309, USA }
\date{\today} 

\begin{abstract}

We study anisotropic thermalization in dilute gases of microwave shielded polar molecular fermions. 
For collision energies above the threshold regime, we find that thermalization is suppressed due to a strong preference for forward scattering and a reduction in total cross section with energy, significantly reducing the efficiency of evaporative cooling. 
We perform close-coupling calculations on the effective potential energy surface derived by Deng et al. [Phys. Rev. Lett. 130, 183001 (2023)], to obtain accurate 2-body elastic differential cross sections across a range of collision energies.  
We use Gaussian process regression to obtain a global representation of the differential cross section, over a wide range of collision angles and energies.
The route to equilibrium is then analyzed with cross-dimensional rethermalization experiments, quantified by a measure of collisional efficiency toward achieving thermalization.

\end{abstract}

\maketitle

The ever growing interest in quantum control of polar molecules motivates the cooling of molecular gases to unprecedented cold temperatures \cite{DeMille02_PRL, Carr09_NJP, Jin12_ACS, Quemener12_ACS, Bohn17_Sci}. In bulk gases, reaching such temperatures can be accomplished through evaporative cooling \cite{Ketterle96_AP}, a process which throws away energetic molecules and leverages collisions to rethermalize the remaining, less energetic, distribution.  
Understanding and controlling 2-body scattering for thermalization is, therefore, of great importance for ultracold experiments. 
To this end, the exciting advent of collisional shielding with external fields has permitted a large suppression of 2-body losses between molecules \cite{Quemener16_PRA, Xie20_PRL, Matsuda20_Sci, Li21_Nat, Mukherjee23_PRR}. Thermalization relies instead on the elastic cross section, which is generally dependent on the field-induced dipole-dipole interaction and their energy of approach. 
Of particular interest to this Letter is collisional shielding with microwave fields \cite{Gorshkov08_PRL, Avdeenkov12_PRA, Lassabliere18_PRL, Karman18_PRL}, recently achieved at several labs around the world \cite{Anderegg21_Sci, Schindewolf22_Nat, Bigagli23_NatPhys, Lin23_PRX}.

In analogous gases of magnetic atoms with comparatively small dipole moments, dipolar scattering remains close-to-threshold \cite{Sadeghpour00_JPB} at the ultracold but nodegenerate temperatures of $T \sim 100$ nK \cite{Aikawa14_PRL, Tang16_PRL, Tang16_PRA, Patscheider21_PRA}. 
For dipoles, threshold scattering occurs when the collision energy is much lower than the dipole energy $E_{\rm dd}$, in which case the scattering cross section becomes energy independent \cite{Bohn09_NJP} with a universal analytic form \cite{Bohn14_PRA}. Numerical studies of thermalization are made much simpler at universality, since collisions can be sampled regardless of collision energy \cite{Sykes15_PRA, Wang21_PRA}.  However, this convenience is lost with the polar molecular gases of interest here.
Take for instance a gas of fermionic $^{23}$Na$^{40}$K, as we will concern ourselves with in this study. This species has a large intrinsic dipole moment of $d = 2.72$ D, so that even ultracold temperatures have majority of collisions occurring away from threshold with an energy dependent cross section. 

In this Letter, we find that non-threshold collisions can dramatically reduce thermalization and thus, the efficiency of the cooling process. 
Ignoring all 1 and 2-body losses for a focused study on elastic collisions, the decrease in gas total energy $E = 3 N k_B T$ along with the number of molecules $N$, approximately follows the coupled rate equations \cite{Luiten96_PRA, Bigagli23_NatPhys}
\begin{subequations} \label{eq:rate_equations}
\begin{align}
    \frac{ d N }{ d t }
    &=
    -  
    \nu(\kappa) 
    \gamma_{\rm th} 
    N, \\
    \frac{ d E }{ d t }
    &=
    -  
    \frac{ 1 }{ 3 }
    \lambda(\kappa)
    \gamma_{\rm th} 
    E,
\end{align}
\end{subequations}
where $\nu(\kappa) = ( 2 + 2\kappa + \kappa^2 ) / ( 2 e^{\kappa} )$ and $\lambda(\kappa) = ( 6 + 6\kappa +3\kappa^2 + \kappa^3 ) / ( 2 e^{\kappa} )$ are functions of the energetic truncation parameter $\kappa = U / (k_B T)$ \cite{Davis95_APB}. 

\begin{figure}[ht]
    \centering
    \includegraphics[width=\columnwidth]{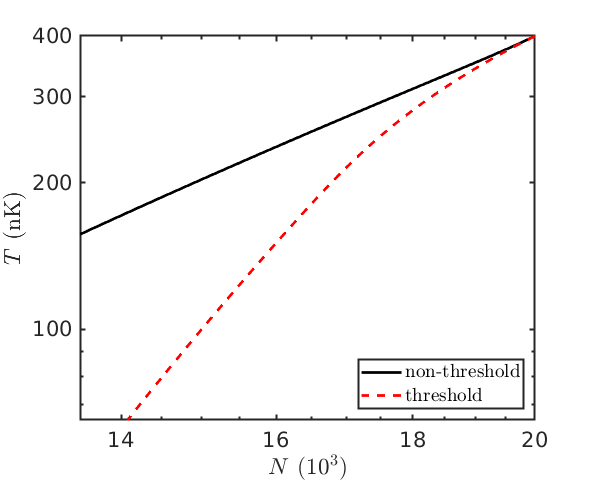}
    \caption{ A log-log plot of $T$ vs $N$ during a forced evaporation protocol. The plot compares the evaporation trajectory for microwave shielded $^{23}$Na$^{40}$K when scattering is realistic and non-threshold (solid black curve), to the artificial case of threshold scattering (dashed red curve). Both 1 and 2-body losses are assumed negligible and ignored here. }
    \label{fig:logT_vs_logN_Theta90}
\end{figure}

By continuously lowering the energetic depth of the confining potential $U(t) = U_0 \exp(-t/\tau)$ over a time interval $\tau$, highly energetic molecules are forced to evaporate away, lowering the number of molecules along with the gas temperature as shown in Fig.~\ref{fig:logT_vs_logN_Theta90}. For the plot, Eq.~(\ref{eq:rate_equations}) is solved by taking evaporation to occur with an initial trap depth $U_0/k_B = 4 \: \mu$K over $\tau = 0.5$ s, in a harmonic trap with mean frequency $\omega = 2\pi \times 100$ Hz, starting at temperature $T_0 = 400$ nK and molecule number $N_0 = 20,000$. 
The evaporation efficiency, defined as the slope of $T$ vs $N$ on a log-log scale, is governed by the thermalization rate $\gamma_{\rm th}$. 
The figure shows efficient cooling for the low-energy threshold cross sections (dashed red curve), and significantly less efficient cooling for the realistic cross sections (solid black curve). 
The remainder of this Letter provides the microscopic mechanisms that lead to this dramatic difference, and efficient theoretical tools we employ to obtain these conclusions.

\begin{figure}[ht]
    \centering
    \includegraphics[width=\columnwidth]{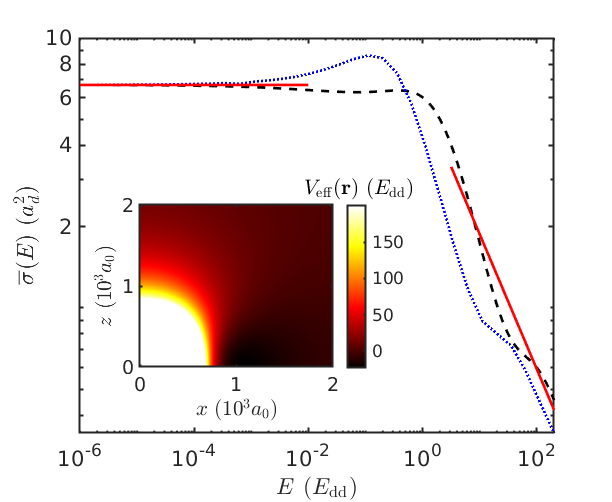}
    \caption{ 
    Energy dependence of the angular averaged total cross section $\overline{\sigma}$ between microwave shielded $^{23}$Na$^{40}$K (black dashed line). 
    The energy dependence clearly differs from the total cross section between fermionic point dipoles (dotted blue curve).
    For comparison, we plot the low energy Born and high energy Eikonal approximations with solid red lines.
    The inset shows a $y = 0$ slice of the effective microwave shielding interaction potential, with Rabi frequency $\Omega = 2\pi \times 15$ MHz and microwave detuning $\Delta = 2 \pi \times 9.5$ MHz. The shielding core is depicted as a white patch surrounding the coordinate origin which saturates the colorbar at $V_{\rm eff} > 200 E_{\rm dd}$. Coordinate axes are plotted in units of $10^3$ Bohr radii $a_0$.
    } 
    \label{fig:sigma_vs_E}
\end{figure}

{\it Shielded collisions}---Central to this study, are collisions that occur between molecules shielded by circularly polarized microwaves \cite{Karman18_PRL}. The resulting potential energy surface between two such molecules is conveniently described by a single effective potential \cite{Deng23_PRL}:
\begin{align} \label{eq:effective_potential}
    V_{\rm eff}(\boldsymbol{{r}})  
    &= 
    \frac{ C_6 }{ r^6 } 
    \big[ 1 - ( \hat{\boldsymbol{r}} \cdot \hat{\boldsymbol{{\cal E}}} )^4 \big] 
    +
    \frac{ \overline{d}^2 }{ 4 \pi \epsilon_0 }
    \frac{ 3 ( \hat{\boldsymbol{r}} \cdot \hat{\boldsymbol{{\cal E}}} )^2 
    -
    1 }{ r^3 },
\end{align}
where $\boldsymbol{r} = ( r, \theta, \phi )$ is the relative position between the two colliding molecules, 
$\hat{\boldsymbol{{\cal E}}}$ is the axis along which the dipoles are effectively aligned, $\overline{d} = d_0 / \sqrt{ 12 (1 + (\Delta/\Omega)^2) }$ is the effective molecular dipole moment and $C_6 = d_0^4 ( 1 +  (\Delta/\Omega)^2)^{-3/2} / ( 128 \pi^2 \epsilon_0^2 \hbar \Omega )$. Here $\Delta$ and $\Omega$ are the detuning and Rabi frequency respectively, of the microwaves.
A $y = 0$ slice of the effective microwave shielding interaction potential is plotted in the inset of Fig.~\ref{fig:sigma_vs_E}.
Notably, the long-range $1/r^3$ tail of $V_{\rm eff}(\boldsymbol{{r}})$ is almost identical to that of point dipole particles, modified only by an overall minus sign. As a result, the close-to-threshold elastic cross sections for microwave shielded molecules are identical to those for point dipoles.

It is natural to introduce units based on the reduced mass $\mu$, dipole length and dipole energy:
\begin{align}
    a_d 
    = 
    \frac{ \mu \overline{d}^2 }{ 4 \pi \epsilon_0 \hbar^2 } 
    \quad\text{and}\quad
    E_{\rm dd}
    =
    \frac{ \hbar^2 }{ \mu a_d^2 }, 
\end{align}
respectively. Threshold scattering is then expected to occur for collision energies $E \ll E_{\rm dd}$. 
With the microwave parameters $\Delta = 2\pi \times 15$ MHz and $\Delta = 2\pi \times 9.5$ MHz, which will be assumed in what follows, the molecules see a dipole length of $a_d \approx 3900 a_0$, corresponding to a dipole energy of $E_{\rm dd}/k_B \approx 360$ nK.  
Therefore, temperatures comparable to $E_{\rm dd} / k_B$ are insufficient to keep molecular scattering in the threshold regime \cite{Bohn09_NJP}. 
Moreover, since the dipole energy scales as $E_{\rm dd} \sim d^{-4}$, larger dipoles require much lower temperatures to achieve universal dipolar threshold scattering as alluded to earlier. 
Away from threshold, the integral cross section $\overline{\sigma}$ in the presence of microwave shielding (dashed black curve), develops a nontrivial energy dependence that clearly differs from that of plain point dipoles (dotted blue curve) as illustrated in Fig.~\ref{fig:sigma_vs_E}. The plotted cross sections were obtained from close-coupling calculations logarithmically spaced in energy, with a universal loss short-range boundary condition \cite{Wang15_NJP} (see Supplementary Material for further details).

Away from threshold at $E \approx E_{\rm dd}$, the microwave shielded integral cross section does not deviate much from its value at threshold (solid red line in Fig.~\ref{fig:sigma_vs_E}). But the differential cross section could still have its anisotropy changed substantially, which is what ultimately affects thermalization \cite{Bohn14_PRA}.
For a study of both non-threshold differential scattering and its implications to thermalization in nondegenerate Fermi gases, we take its nonequilibrium evolution as governed the Boltzmann transport equation \cite{Pitaevskii17_ES}.  
Formulated in this way, numerical solutions treat the molecular positions and momenta as classical variables, while collisions can be efficiently computed by means of Monte Carlo sampling \cite{Bird70_PF, Sykes15_PRA}.  
But on the fly close-coupling calculations would be too expensive for such sampling over a broad range of collision energies and angles. Instead, we propose the following.

{\it Gaussian process fitting}---At a given collision energy, the  elastic differential cross section ${\cal D}_{\rm el}$, is a function of the dipole alignment axis $\hat{\boldsymbol{{\cal E}}}$, and the relative ingoing and outgoing momentum vectors $\hbar \boldsymbol{k}$ and $\hbar \boldsymbol{k}'$, respectively. 
Collectively, we refer to this set of parameters as $\boldsymbol{\beta}$. 
By first performing close-coupling calculations at several well chosen collision energies $E = \hbar^2 k^2 / (2 \mu)$ \footnote{
The differential cross section suffers innate convergence issues due to singularities in the scattering amplitude \cite{Bohn14_PRA}. Fortunately for us, forward scattering does not contribute toward cross-dimensional thermalization, of which we are concerned with in this Letter. We leave addressing these issues to a future manuscript.  
}, we can use the resultant scattering data to infer an $M$-dimensional continuous hypersurface that approximates ${\cal D}_{\rm el}$, with a Gaussian process (GP) model \cite{Sacks89_TF, Cui16_JPB, Christianen19_JCP}.

GP regression is a machine learning technique used to interpolate discrete data points, stitching them together to form a continuous global surface. To do so, a GP assumes that ${\cal D}_{\rm el}(\boldsymbol{\beta})$ evaluated any 2 nearby points in its coordinate space, $\boldsymbol{\beta}_i$ and $\boldsymbol{\beta}_j$, are Gaussian distributed with a covariance given in terms of a function $K(\boldsymbol{\beta}_i, \boldsymbol{\beta}_j)$, called the kernel. A parameterized functional form for the kernel is chosen prior to the surface fitting process, reducing the task of combing through an infinite space of possible functions that best match the data, to a minimization over the kernel parameters.
This minimization step is referred to as \textit{training} the GP model.

\onecolumngrid

\begin{figure}[ht]
    \centering
    \includegraphics[width=\columnwidth]{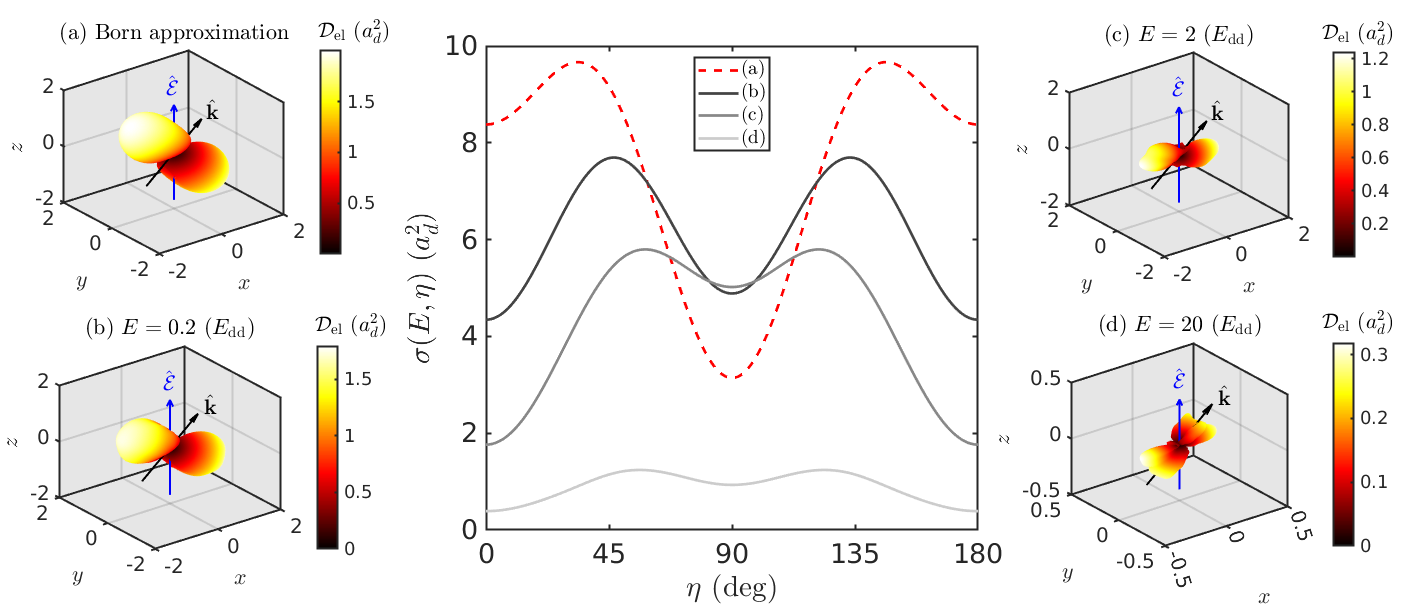}
    \caption{ 
    The central plot shows the total cross section as a function of the incident collision angle, obtained from (a) the Born approximation (red dashed curve), and from GP interpolation (solid curves) for 3 different collision energies: (b) $E = 0.2 E_{\rm dd}$ (black), (c) $E = 2 E_{\rm dd}$ (gray) and (d) $E = 20 E_{\rm dd}$ (light gray). In alphabetical correspondence, are angular plots of the differential cross section (in units of $a_d^2$) in subplots with the respective collision energies, assuming dipoles pointing along $\hat{\boldsymbol{{\cal E}}} = \hat{\boldsymbol{z}}$ and incident collision angle $\eta = 45^{\circ}$ lying in the $x,z$-plane. 
    Subplot (d) uses a smaller domain for clarity of presentation.  
    }
    \label{fig:TCSnDCS_vs_Eeta}
\end{figure}

\twocolumngrid

Several symmetries in the differential cross section help to reduce the computational load of training slightly.  
Rotated into the frame where $\hat{\boldsymbol{{\cal E}}}$ points along the $z$ axis, which we refer to as the dipole-frame, the unique hypersurface regions effectively live in an $M = 4$ dimensional space, with coordinates $\boldsymbol{\beta} = (E, \eta, \theta_s, \phi_s)$. 
As defined, $\eta = \cos^{-1} \hat{\boldsymbol{k}} \cdot \hat{\boldsymbol{{\cal E}}}$ is the angle between the dipole and incident relative momentum directions, where it is convenient to select $\hat{\boldsymbol{k}}$ to lie in its $x,z$ plane.  The angles $\theta_s$ and $\phi_s$, denote the inclination and azimuthal scattering angles respectively, in this frame. Doing so, the differential cross section possesses the symmetry
\begin{align}
    {\cal D}_{\rm el}
    (E, \eta, \theta_s, \phi_s)
    &=
    {\cal D}_{\rm el}
    (E, \eta, \theta_s, -\phi_s). 
\end{align}
Consequently, we only need to specify the differential cross section for angles within the domain $\eta, \theta_s, \phi_s \in [ 0, \pi ]$, to fully describe its global structure.
More details of the appropriate frame transformations are provided in Supplementary Material.

To perform the interpolation with GP regression, we utilize the Mat\'ern-$\frac{5}{2}$ kernel \cite{Rasmussen05_MIT}, 
which is better able to capture the sharp jumps in a non-smooth function, over higher-order differentiable kernels such as the radial basis function. This kernel contains a parameter $w$ that sets a length scale over which features of the data vary in coordinate space, that is optimized during the model training process. 
This kernel is typically not ideal for periodic input data, so we make the periodicity of the angles $( \eta, \theta_s, \phi_s)$ explicitly known to the GP model by training it with the cosine of these angles, instead of the angles themselves. 
Furthermore, $\log_{10}(E / E_{\rm dd})$ is fed into the GP model in place of $E$, to reduce the disparity in fitting domains between each coordinate of $\boldsymbol{\beta}$. The GP  model is trained over the range $\log_{10}(E / E_{\rm dd}) = -6$ to 2, corresponding to collision energies of $E / k_B \approx 0.36$ pK to $36$ $\mu$K.  
After training on $\sim 10,000$ samples of ${\cal D}_{\rm el}(E, \eta, \theta_s, \phi_s)$, the resulting GP fit obtains a mean-squared error of $\approx 0.5 \%$ against the close-coupling calculations 
\footnote{ We utilize more points than is usually necessary for GP fitting in this study, so as to obtain more accurate results of subsequently computed quantities in this Letter. We also optimize the model's hyperparameters \cite{Yang20_NC} on top of just the kernel parameters.  
Even so, the Gaussian process model has issues faithfully reproducing the differential cross section around $\eta, \theta_s = 90^{\circ}$, known to have a discontinuity at threshold \cite{Bohn14_PRA}. Fortunately, this angular segment corresponds to forward scattering, which does not contribute to the cross-dimensional thermalization process of interest here. We ignore this issue until necessary for consideration in future works.      
}, 
which we take as an accurate representation of the actual cross section.

In Fig.~\ref{fig:TCSnDCS_vs_Eeta}, we plot the total cross section $\sigma(E, \eta) = \int {\cal D}_{\rm el}(E, \eta, \Omega_s) d\Omega_s$, at various collision energies. 
There is a marked variation in the $\eta$ dependence, indicating a higher tendency for side-to-side collisions ($\eta=90^{\circ}$) over head-to-tail ones ($\eta=0^{\circ}$) at higher energies.  
To highlight the dominant anisotropic scattering process, Fig.~\ref{fig:TCSnDCS_vs_Eeta} also provides plots of the differential cross section at $\eta = 45^{\circ}$, the approximate angle at which $\sigma$ is maximal.  
As energy increases from subplots (a) to (d), the scattered angle dependence of ${\cal D}_{\rm el}$ becomes biased toward forward scattering, reducing the effectiveness of collisions for thermalization as discussed below.  
Alphabetic labels in Fig.~\ref{fig:TCSnDCS_vs_Eeta} consistently correspond to the collision energies: (b) $E = 0.2 E_{\rm dd}$, (c) $E = 2 E_{\rm dd}$ and (d) $E = 20 E_{\rm dd}$.  
The Born approximated cross sections at threshold \cite{Bohn14_PRA} are labeled with (a).

{\it Collisional thermalization}---Fast and easy access to the accurate differential cross section via its GP model now permits accurate theoretical investigations of nondegenerate gas dynamics. 
More specifically, we are concerned here with a gas' route to thermal equilibrium.  
A common experiment for such analysis is cross-dimensional rethermalization \cite{Monroe93_PRL}, in which a harmonically trapped gas is excited along one axis, then left alone to re-equilibrate from collisions.

We present results in terms of the temperatures along each axis $i$, defined in the presence of a harmonic trap as ${\cal T}_i = ( { \langle p_i^2 \rangle / m } + m \omega_i^2 \langle q_i^2 \rangle ) / 2$,
where $\langle \ldots \rangle = \int d^3 \boldsymbol{q} d^3 \boldsymbol{p} f(\boldsymbol{q}, \boldsymbol{p}) ( \ldots )$ denotes a phase space average over the phase space distribution $f$ in molecular positions $\boldsymbol{q}$ and momenta $\boldsymbol{p}$, while 
$\omega_i$ are the harmonic trapping frequencies. 
As is usual in cross-dimensional rethermalization, we consider an excitation of axis $i$ then proceed to measure the thermalization rate along axis $j$. This is modeled by taking axis $i$ to have an initial out-of-equilibrium temperature ${\cal T}_i = T_{0} + {\delta_i/k_B}$,
with a perturbance in energy $\delta_i$, while the the other 2 axes are simply at initial temperature $T_0$.

\onecolumngrid

\begin{figure}[ht]
    \centering
    \includegraphics[width=\columnwidth]{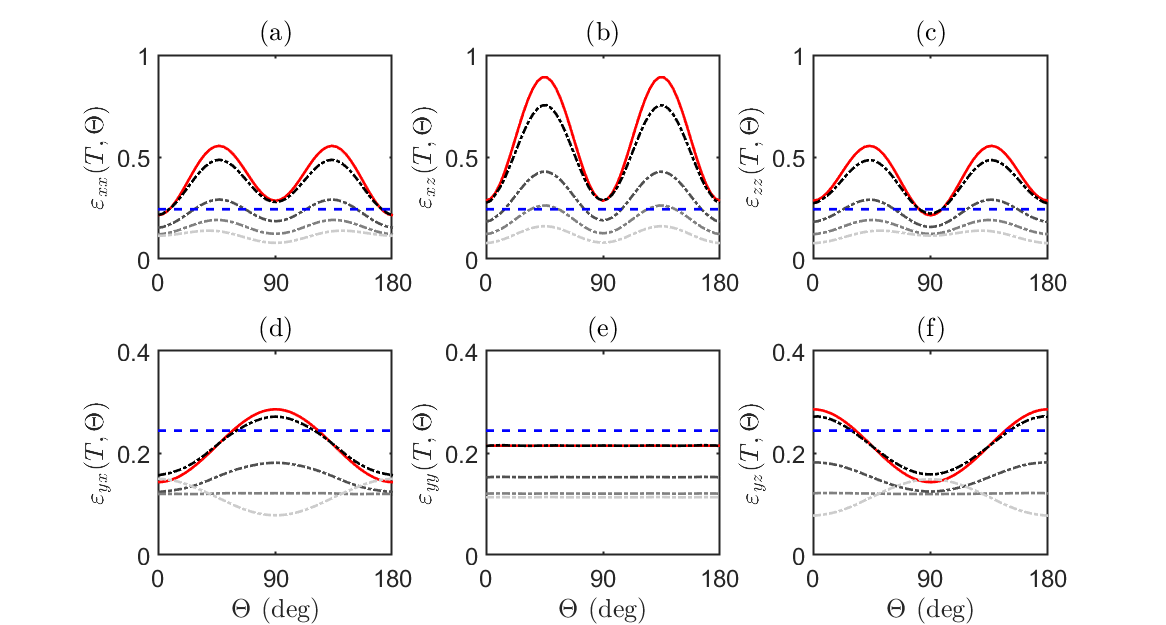} 
    \caption{ Plot of $\varepsilon_{i j}$ as a function of the dipole tilt angle $\Theta$, for all 6 unique configurations (subplots a to f) of the excitation axis $i$, and measured thermalization axis $j$. The solid red curves are the analytic $\varepsilon_{i j}$ results derived with the Born approximated cross section at threshold, whereas the dashed-dotted curves are those from Monte Carlo integration using the GP interpolated cross sections, at temperatures $T = 10$ nK (black), $T = 100$ nK (dark gray), $T = 400$ nK (gray), and $T = 1$ $\mu$K (light gray). The dashed blue lines are the efficiency for purely $p$-wave collisions, $\varepsilon_p = 1/4.1$. }
    \label{fig:NCPR_GPvsBorn}
\end{figure}

\twocolumngrid

In the case of a dilute gas, the relaxation of ${\cal T}_j$ follows an exponential decay in time, whose rate $\gamma_{ij}$ is related to the standard collision rate $\gamma_{\rm coll}$,
by a proportionality factor $\varepsilon_{ij} = { \gamma_{ij} / \gamma_{\rm coll} }$.
As defined, the quantity $\varepsilon_{i j}$ is the inverse of the so-called number of collisions per rethermalization \cite{Snoke89_PRB, Monroe93_PRL}, a measure of thermalization common to the literature \cite{Li21_Nat, Schindewolf22_Nat, Bigagli23_NatPhys}. We opt to utilize its inverse instead as it is the more natural definition to discuss efficiency of evaporative cooling.
Usually defined as $\gamma_{\rm coll} = \langle n \rangle \langle \overline{\sigma} v_r \rangle$ with phase space averaged number density $\langle n \rangle$ and 2-body elastic rate $\langle \overline{\sigma} v_r \rangle$, $\varepsilon_{ij}$ represents the efficiency of each non-threshold collision toward thermalization of the gas. 
This collisional efficiency is formally cast in terms of the integral
\begin{align} \label{eq:NCPR_integral}
    \varepsilon_{ij} 
    &\approx
    \alpha_{i j} 
    \frac{ \pi^2 }{ 64 }
    \int \frac{ d^3 \boldsymbol{\kappa}  }{ ( 2 \pi )^3 }
    \frac{ e^{ -{ \kappa^2 / 4 } } }{ \sqrt{ \pi } }
    \int d^2\Omega' 
    \frac{
    {\cal D}'_{\rm el} \kappa
    }{
    \langle \sigma \kappa \rangle 
    }
    \Delta \kappa^2_{i}
    \Delta \kappa^2_{j},
\end{align}
where $\Delta \kappa_{i}^2 = \kappa_{i}'^2 - \kappa_{i}^2$ is the collisional change in adimensional relative momenta $\boldsymbol{\kappa} = \boldsymbol{p}_{r} ( m k_B T_0 )^{-1/2}$, 
$\alpha_{i j} = 3/2$ if $i = j$, and $\alpha_{i j} = -3$ otherwise (see Supplementary Materials). 
The integral above has been evaluated analytically in the threshold scattering regime \cite{Wang21_PRA}, both for identical dipolar fermions and bosons.

Evidently from Eq.~(\ref{eq:NCPR_integral}), $\varepsilon_{i j}$ is symmetric in its indices which leaves only 6 unique configurations of $i$ and $j$. 
Asserting the dipoles lie in the $x,z$-plane and tilted with angle $\Theta = \cos^{-1} \hat{\boldsymbol{{\cal E}}} \cdot \hat{\boldsymbol{z}}$,  
we compute Eq.~(\ref{eq:NCPR_integral}) with Monte Carlo integration \footnote{
The Monte Carlo integration gives a $\lesssim 1\%$ error, which is mostly imperceptible in the log-linear plot. 
} and plot the results in Fig.~\ref{fig:NCPR_GPvsBorn}. Each subplot (a to f) shows a different ($i, j$) configuration, within which, $\varepsilon_{i j}$ is plotted against the dipole tilt angle $\Theta$ as dashed curves, for the temperatures $T = 10$ nK (black), $T = 100$ nK (dark gray), $T = 400$ nK (gray) and $T = 1$ $\mu$K (light gray). 
Interestingly, the $\varepsilon_{i j}$ terms involving excitation or rethermalization along $y$ essentially lose their dependence on $\Theta$ around $400$ nK, beyond which collisions are less efficient than even nondipolar $p$-wave scattering (dashed blue line in Fig.~\ref{fig:NCPR_GPvsBorn}) \cite{DeMarco99_PRL} for all $\Theta$.   
This decrease can be intuited by looking at the differential cross section around $\eta = 45^{\circ}$, around which the total cross section is maximal.
As evidenced from the subplots of ${\cal D}_{\rm el}$ in Fig.~\ref{fig:TCSnDCS_vs_Eeta}, forward scattering is favored at higher collision energies, limiting momentum transfer between axes and therefore, also the efficiency of collisions toward rethermalization.  
Preferential forward scattering is what ultimately leads to the reduction in evaporation efficiency, earlier described and seen in Fig.~\ref{fig:logT_vs_logN_Theta90}. There, the rate of thermalization was approximated by the average $\gamma_{\rm th} = \gamma_{\rm coll} \sum_{i,j} \varepsilon_{i j} / 9$, as is expected for evaporation along all 3-dimensions. The dipoles were assumed aligned along $\Theta = 90^{\circ}$, and $\gamma_{\rm th}$ interpolated over several temperatures to solve Eq.~(\ref{eq:rate_equations}).

Realistically, forced evaporation by trap depth lowering tends to occur primarily along 1 direction, reducing the evaporation efficiency in the presence of molecular losses \cite{Surkov96_PRA}. The resulting out-of-equilibrium momentum distribution from single axis evaporation will be much like that in cross-dimensional rethermalization experiments, where an anisotropic collisional efficiency could now be used to your advantage. For instance, near unity collisional efficiency is achieved in the threshold regime with $\varepsilon_{x z}$ specifically at $\Theta = 45^{\circ}$. Optimal evaporation protocols could thus be engineered by varying the molecular dipole orientation relative to the axis of evaporation. We leave such investigations to a future work.

{\it Outlook and conclusions}---By constructing a GP model of the elastic differential cross section between microwave shielded polar molecular fermions, we have found that non-threshold collisions can greatly diminish the efficacy of collisions toward thermalization of a nondegenerate gas. 
It is thus prudent to perform evaporation in the threshold regime, with the caveat that Pauli blocking in fermions would also lower the collisional efficiency below the Fermi temperature \cite{Aikawa14_PRL}. 
If deployed in direct simulation Monte Carlo solvers \cite{Bird70_PF, Sykes15_PRA, Wang21_PRA}, this GP model could also permit accurate dynamical studies in the Fermi degenerate or hydrodynamic regimes.   
The latter is motivated by restrictions of $\varepsilon_{i j}$, only being able to describe thermalization in dilute samples. With larger molecular dipoles at densities required to achieve quantum degeneracy, the collision rate is far exceeded by the mean trapping frequency, demanding equilibration of trapped dipolar gases be treated within a hydrodynamic framework \cite{Wang22_PRA, Wang22_PRA2, Wang23_PRA, Wang23_PRA2}. 
The method of GP interpolation proposed here could similarly be applied to DC field shielded molecules \cite{Lassabliere22_PRA} and bosonic species.

{\it Acknowledgments}---The authors are grateful to Luo Xin-Yu for motivating discussions and insights on evaporation in molecular Fermi gases.
This work is supported by the National Science Foundation under Grant Number PHY2110327.

\bibliography{main.bib} 

\begin{thebibliography}{56}%
\makeatletter
\providecommand \@ifxundefined [1]{%
 \@ifx{#1\undefined}
}%
\providecommand \@ifnum [1]{%
 \ifnum #1\expandafter \@firstoftwo
 \else \expandafter \@secondoftwo
 \fi
}%
\providecommand \@ifx [1]{%
 \ifx #1\expandafter \@firstoftwo
 \else \expandafter \@secondoftwo
 \fi
}%
\providecommand \natexlab [1]{#1}%
\providecommand \enquote  [1]{``#1''}%
\providecommand \bibnamefont  [1]{#1}%
\providecommand \bibfnamefont [1]{#1}%
\providecommand \citenamefont [1]{#1}%
\providecommand \href@noop [0]{\@secondoftwo}%
\providecommand \href [0]{\begingroup \@sanitize@url \@href}%
\providecommand \@href[1]{\@@startlink{#1}\@@href}%
\providecommand \@@href[1]{\endgroup#1\@@endlink}%
\providecommand \@sanitize@url [0]{\catcode `\\12\catcode `\$12\catcode
  `\&12\catcode `\#12\catcode `\^12\catcode `\_12\catcode `\%12\relax}%
\providecommand \@@startlink[1]{}%
\providecommand \@@endlink[0]{}%
\providecommand \url  [0]{\begingroup\@sanitize@url \@url }%
\providecommand \@url [1]{\endgroup\@href {#1}{\urlprefix }}%
\providecommand \urlprefix  [0]{URL }%
\providecommand \Eprint [0]{\href }%
\providecommand \doibase [0]{https://doi.org/}%
\providecommand \selectlanguage [0]{\@gobble}%
\providecommand \bibinfo  [0]{\@secondoftwo}%
\providecommand \bibfield  [0]{\@secondoftwo}%
\providecommand \translation [1]{[#1]}%
\providecommand \BibitemOpen [0]{}%
\providecommand \bibitemStop [0]{}%
\providecommand \bibitemNoStop [0]{.\EOS\space}%
\providecommand \EOS [0]{\spacefactor3000\relax}%
\providecommand \BibitemShut  [1]{\csname bibitem#1\endcsname}%
\let\auto@bib@innerbib\@empty
\bibitem [{\citenamefont {DeMille}(2002)}]{DeMille02_PRL}%
  \BibitemOpen
  \bibfield  {author} {\bibinfo {author} {\bibfnamefont {D.}~\bibnamefont
  {DeMille}},\ }\href {https://doi.org/10.1103/PhysRevLett.88.067901}
  {\bibfield  {journal} {\bibinfo  {journal} {Phys. Rev. Lett.}\ }\textbf
  {\bibinfo {volume} {88}},\ \bibinfo {pages} {067901} (\bibinfo {year}
  {2002})}\BibitemShut {NoStop}%
\bibitem [{\citenamefont {Carr}\ \emph {et~al.}(2009)\citenamefont {Carr},
  \citenamefont {DeMille}, \citenamefont {Krems},\ and\ \citenamefont
  {Ye}}]{Carr09_NJP}%
  \BibitemOpen
  \bibfield  {author} {\bibinfo {author} {\bibfnamefont {L.~D.}\ \bibnamefont
  {Carr}}, \bibinfo {author} {\bibfnamefont {D.}~\bibnamefont {DeMille}},
  \bibinfo {author} {\bibfnamefont {R.~V.}\ \bibnamefont {Krems}},\ and\
  \bibinfo {author} {\bibfnamefont {J.}~\bibnamefont {Ye}},\ }\href
  {https://doi.org/10.1088/1367-2630/11/5/055049} {\bibfield  {journal}
  {\bibinfo  {journal} {New Journal of Physics}\ }\textbf {\bibinfo {volume}
  {11}},\ \bibinfo {pages} {055049} (\bibinfo {year} {2009})}\BibitemShut
  {NoStop}%
\bibitem [{\citenamefont {Jin}\ and\ \citenamefont {Ye}(2012)}]{Jin12_ACS}%
  \BibitemOpen
  \bibfield  {author} {\bibinfo {author} {\bibfnamefont {D.~S.}\ \bibnamefont
  {Jin}}\ and\ \bibinfo {author} {\bibfnamefont {J.}~\bibnamefont {Ye}},\
  }\href {https://doi.org/10.1021/cr300342x} {\bibfield  {journal} {\bibinfo
  {journal} {Chemical Reviews}\ }\textbf {\bibinfo {volume} {112}},\ \bibinfo
  {pages} {4801} (\bibinfo {year} {2012})}\BibitemShut {NoStop}%
\bibitem [{\citenamefont {Qu{\'e}m{\'e}ner}\ and\ \citenamefont
  {Julienne}(2012)}]{Quemener12_ACS}%
  \BibitemOpen
  \bibfield  {author} {\bibinfo {author} {\bibfnamefont {G.}~\bibnamefont
  {Qu{\'e}m{\'e}ner}}\ and\ \bibinfo {author} {\bibfnamefont {P.~S.}\
  \bibnamefont {Julienne}},\ }\href {https://doi.org/10.1021/cr300092g}
  {\bibfield  {journal} {\bibinfo  {journal} {Chemical Reviews}\ }\textbf
  {\bibinfo {volume} {112}},\ \bibinfo {pages} {4949} (\bibinfo {year}
  {2012})}\BibitemShut {NoStop}%
\bibitem [{\citenamefont {Bohn}\ \emph {et~al.}(2017)\citenamefont {Bohn},
  \citenamefont {Rey},\ and\ \citenamefont {Ye}}]{Bohn17_Sci}%
  \BibitemOpen
  \bibfield  {author} {\bibinfo {author} {\bibfnamefont {J.~L.}\ \bibnamefont
  {Bohn}}, \bibinfo {author} {\bibfnamefont {A.~M.}\ \bibnamefont {Rey}},\ and\
  \bibinfo {author} {\bibfnamefont {J.}~\bibnamefont {Ye}},\ }\href
  {https://doi.org/10.1126/science.aam6299} {\bibfield  {journal} {\bibinfo
  {journal} {Science}\ }\textbf {\bibinfo {volume} {357}},\ \bibinfo {pages}
  {1002} (\bibinfo {year} {2017})}\BibitemShut {NoStop}%
\bibitem [{\citenamefont {Ketterle}\ and\ \citenamefont
  {Druten}(1996)}]{Ketterle96_AP}%
  \BibitemOpen
  \bibfield  {author} {\bibinfo {author} {\bibfnamefont {W.}~\bibnamefont
  {Ketterle}}\ and\ \bibinfo {author} {\bibfnamefont {N.~V.}\ \bibnamefont
  {Druten}}\ }(\bibinfo  {publisher} {Academic Press},\ \bibinfo {year}
  {1996})\BibitemShut {NoStop}%
\bibitem [{\citenamefont {Qu\'em\'ener}\ and\ \citenamefont
  {Bohn}(2016)}]{Quemener16_PRA}%
  \BibitemOpen
  \bibfield  {author} {\bibinfo {author} {\bibfnamefont {G.}~\bibnamefont
  {Qu\'em\'ener}}\ and\ \bibinfo {author} {\bibfnamefont {J.~L.}\ \bibnamefont
  {Bohn}},\ }\href {https://doi.org/10.1103/PhysRevA.93.012704} {\bibfield
  {journal} {\bibinfo  {journal} {Phys. Rev. A}\ }\textbf {\bibinfo {volume}
  {93}},\ \bibinfo {pages} {012704} (\bibinfo {year} {2016})}\BibitemShut
  {NoStop}%
\bibitem [{\citenamefont {Xie}\ \emph {et~al.}(2020)\citenamefont {Xie},
  \citenamefont {Lepers}, \citenamefont {Vexiau}, \citenamefont {Orb\'an},
  \citenamefont {Dulieu},\ and\ \citenamefont {Bouloufa-Maafa}}]{Xie20_PRL}%
  \BibitemOpen
  \bibfield  {author} {\bibinfo {author} {\bibfnamefont {T.}~\bibnamefont
  {Xie}}, \bibinfo {author} {\bibfnamefont {M.}~\bibnamefont {Lepers}},
  \bibinfo {author} {\bibfnamefont {R.}~\bibnamefont {Vexiau}}, \bibinfo
  {author} {\bibfnamefont {A.}~\bibnamefont {Orb\'an}}, \bibinfo {author}
  {\bibfnamefont {O.}~\bibnamefont {Dulieu}},\ and\ \bibinfo {author}
  {\bibfnamefont {N.}~\bibnamefont {Bouloufa-Maafa}},\ }\href
  {https://doi.org/10.1103/PhysRevLett.125.153202} {\bibfield  {journal}
  {\bibinfo  {journal} {Phys. Rev. Lett.}\ }\textbf {\bibinfo {volume} {125}},\
  \bibinfo {pages} {153202} (\bibinfo {year} {2020})}\BibitemShut {NoStop}%
\bibitem [{\citenamefont {Matsuda}\ \emph {et~al.}(2020)\citenamefont
  {Matsuda}, \citenamefont {Marco}, \citenamefont {Li}, \citenamefont {Tobias},
  \citenamefont {Valtolina}, \citenamefont {Quéméner},\ and\ \citenamefont
  {Ye}}]{Matsuda20_Sci}%
  \BibitemOpen
  \bibfield  {author} {\bibinfo {author} {\bibfnamefont {K.}~\bibnamefont
  {Matsuda}}, \bibinfo {author} {\bibfnamefont {L.~D.}\ \bibnamefont {Marco}},
  \bibinfo {author} {\bibfnamefont {J.-R.}\ \bibnamefont {Li}}, \bibinfo
  {author} {\bibfnamefont {W.~G.}\ \bibnamefont {Tobias}}, \bibinfo {author}
  {\bibfnamefont {G.}~\bibnamefont {Valtolina}}, \bibinfo {author}
  {\bibfnamefont {G.}~\bibnamefont {Quéméner}},\ and\ \bibinfo {author}
  {\bibfnamefont {J.}~\bibnamefont {Ye}},\ }\href
  {https://doi.org/10.1126/science.abe7370} {\bibfield  {journal} {\bibinfo
  {journal} {Science}\ }\textbf {\bibinfo {volume} {370}},\ \bibinfo {pages}
  {1324} (\bibinfo {year} {2020})}\BibitemShut {NoStop}%
\bibitem [{\citenamefont {Li}\ \emph {et~al.}(2021)\citenamefont {Li},
  \citenamefont {Tobias}, \citenamefont {Matsuda}, \citenamefont {Miller},
  \citenamefont {Valtolina}, \citenamefont {De~Marco}, \citenamefont {Wang},
  \citenamefont {Lassabli{\`e}re}, \citenamefont {Qu{\'e}m{\'e}ner},
  \citenamefont {Bohn} \emph {et~al.}}]{Li21_Nat}%
  \BibitemOpen
  \bibfield  {author} {\bibinfo {author} {\bibfnamefont {J.-R.}\ \bibnamefont
  {Li}}, \bibinfo {author} {\bibfnamefont {W.~G.}\ \bibnamefont {Tobias}},
  \bibinfo {author} {\bibfnamefont {K.}~\bibnamefont {Matsuda}}, \bibinfo
  {author} {\bibfnamefont {C.}~\bibnamefont {Miller}}, \bibinfo {author}
  {\bibfnamefont {G.}~\bibnamefont {Valtolina}}, \bibinfo {author}
  {\bibfnamefont {L.}~\bibnamefont {De~Marco}}, \bibinfo {author}
  {\bibfnamefont {R.~R.}\ \bibnamefont {Wang}}, \bibinfo {author}
  {\bibfnamefont {L.}~\bibnamefont {Lassabli{\`e}re}}, \bibinfo {author}
  {\bibfnamefont {G.}~\bibnamefont {Qu{\'e}m{\'e}ner}}, \bibinfo {author}
  {\bibfnamefont {J.~L.}\ \bibnamefont {Bohn}}, \emph {et~al.},\ }\href
  {https://www.nature.com/articles/s41567-021-01329-6} {\bibfield  {journal}
  {\bibinfo  {journal} {Nature Physics}\ }\textbf {\bibinfo {volume} {17}},\
  \bibinfo {pages} {1144} (\bibinfo {year} {2021})}\BibitemShut {NoStop}%
\bibitem [{\citenamefont {Mukherjee}\ \emph {et~al.}(2023)\citenamefont
  {Mukherjee}, \citenamefont {Frye}, \citenamefont {Le~Sueur}, \citenamefont
  {Tarbutt},\ and\ \citenamefont {Hutson}}]{Mukherjee23_PRR}%
  \BibitemOpen
  \bibfield  {author} {\bibinfo {author} {\bibfnamefont {B.}~\bibnamefont
  {Mukherjee}}, \bibinfo {author} {\bibfnamefont {M.~D.}\ \bibnamefont {Frye}},
  \bibinfo {author} {\bibfnamefont {C.~R.}\ \bibnamefont {Le~Sueur}}, \bibinfo
  {author} {\bibfnamefont {M.~R.}\ \bibnamefont {Tarbutt}},\ and\ \bibinfo
  {author} {\bibfnamefont {J.~M.}\ \bibnamefont {Hutson}},\ }\href
  {https://doi.org/10.1103/PhysRevResearch.5.033097} {\bibfield  {journal}
  {\bibinfo  {journal} {Phys. Rev. Res.}\ }\textbf {\bibinfo {volume} {5}},\
  \bibinfo {pages} {033097} (\bibinfo {year} {2023})}\BibitemShut {NoStop}%
\bibitem [{\citenamefont {Gorshkov}\ \emph {et~al.}(2008)\citenamefont
  {Gorshkov}, \citenamefont {Rabl}, \citenamefont {Pupillo}, \citenamefont
  {Micheli}, \citenamefont {Zoller}, \citenamefont {Lukin},\ and\ \citenamefont
  {B\"uchler}}]{Gorshkov08_PRL}%
  \BibitemOpen
  \bibfield  {author} {\bibinfo {author} {\bibfnamefont {A.~V.}\ \bibnamefont
  {Gorshkov}}, \bibinfo {author} {\bibfnamefont {P.}~\bibnamefont {Rabl}},
  \bibinfo {author} {\bibfnamefont {G.}~\bibnamefont {Pupillo}}, \bibinfo
  {author} {\bibfnamefont {A.}~\bibnamefont {Micheli}}, \bibinfo {author}
  {\bibfnamefont {P.}~\bibnamefont {Zoller}}, \bibinfo {author} {\bibfnamefont
  {M.~D.}\ \bibnamefont {Lukin}},\ and\ \bibinfo {author} {\bibfnamefont
  {H.~P.}\ \bibnamefont {B\"uchler}},\ }\href
  {https://doi.org/10.1103/PhysRevLett.101.073201} {\bibfield  {journal}
  {\bibinfo  {journal} {Phys. Rev. Lett.}\ }\textbf {\bibinfo {volume} {101}},\
  \bibinfo {pages} {073201} (\bibinfo {year} {2008})}\BibitemShut {NoStop}%
\bibitem [{\citenamefont {Avdeenkov}(2012)}]{Avdeenkov12_PRA}%
  \BibitemOpen
  \bibfield  {author} {\bibinfo {author} {\bibfnamefont {A.~V.}\ \bibnamefont
  {Avdeenkov}},\ }\href {https://doi.org/10.1103/PhysRevA.86.022707} {\bibfield
   {journal} {\bibinfo  {journal} {Phys. Rev. A}\ }\textbf {\bibinfo {volume}
  {86}},\ \bibinfo {pages} {022707} (\bibinfo {year} {2012})}\BibitemShut
  {NoStop}%
\bibitem [{\citenamefont {Lassabli\`ere}\ and\ \citenamefont
  {Qu\'em\'ener}(2018)}]{Lassabliere18_PRL}%
  \BibitemOpen
  \bibfield  {author} {\bibinfo {author} {\bibfnamefont {L.}~\bibnamefont
  {Lassabli\`ere}}\ and\ \bibinfo {author} {\bibfnamefont {G.}~\bibnamefont
  {Qu\'em\'ener}},\ }\href {https://doi.org/10.1103/PhysRevLett.121.163402}
  {\bibfield  {journal} {\bibinfo  {journal} {Phys. Rev. Lett.}\ }\textbf
  {\bibinfo {volume} {121}},\ \bibinfo {pages} {163402} (\bibinfo {year}
  {2018})}\BibitemShut {NoStop}%
\bibitem [{\citenamefont {Karman}\ and\ \citenamefont
  {Hutson}(2018)}]{Karman18_PRL}%
  \BibitemOpen
  \bibfield  {author} {\bibinfo {author} {\bibfnamefont {T.}~\bibnamefont
  {Karman}}\ and\ \bibinfo {author} {\bibfnamefont {J.~M.}\ \bibnamefont
  {Hutson}},\ }\href {https://doi.org/10.1103/PhysRevLett.121.163401}
  {\bibfield  {journal} {\bibinfo  {journal} {Phys. Rev. Lett.}\ }\textbf
  {\bibinfo {volume} {121}},\ \bibinfo {pages} {163401} (\bibinfo {year}
  {2018})}\BibitemShut {NoStop}%
\bibitem [{\citenamefont {Anderegg}\ \emph {et~al.}(2021)\citenamefont
  {Anderegg}, \citenamefont {Burchesky}, \citenamefont {Bao}, \citenamefont
  {Yu}, \citenamefont {Karman}, \citenamefont {Chae}, \citenamefont {Ni},
  \citenamefont {Ketterle},\ and\ \citenamefont {Doyle}}]{Anderegg21_Sci}%
  \BibitemOpen
  \bibfield  {author} {\bibinfo {author} {\bibfnamefont {L.}~\bibnamefont
  {Anderegg}}, \bibinfo {author} {\bibfnamefont {S.}~\bibnamefont {Burchesky}},
  \bibinfo {author} {\bibfnamefont {Y.}~\bibnamefont {Bao}}, \bibinfo {author}
  {\bibfnamefont {S.~S.}\ \bibnamefont {Yu}}, \bibinfo {author} {\bibfnamefont
  {T.}~\bibnamefont {Karman}}, \bibinfo {author} {\bibfnamefont
  {E.}~\bibnamefont {Chae}}, \bibinfo {author} {\bibfnamefont {K.-K.}\
  \bibnamefont {Ni}}, \bibinfo {author} {\bibfnamefont {W.}~\bibnamefont
  {Ketterle}},\ and\ \bibinfo {author} {\bibfnamefont {J.~M.}\ \bibnamefont
  {Doyle}},\ }\href {https://doi.org/10.1126/science.abg9502} {\bibfield
  {journal} {\bibinfo  {journal} {Science}\ }\textbf {\bibinfo {volume}
  {373}},\ \bibinfo {pages} {779} (\bibinfo {year} {2021})}\BibitemShut
  {NoStop}%
\bibitem [{\citenamefont {Schindewolf}\ \emph {et~al.}(2022)\citenamefont
  {Schindewolf}, \citenamefont {Bause}, \citenamefont {Chen}, \citenamefont
  {Duda}, \citenamefont {Karman}, \citenamefont {Bloch},\ and\ \citenamefont
  {Luo}}]{Schindewolf22_Nat}%
  \BibitemOpen
  \bibfield  {author} {\bibinfo {author} {\bibfnamefont {A.}~\bibnamefont
  {Schindewolf}}, \bibinfo {author} {\bibfnamefont {R.}~\bibnamefont {Bause}},
  \bibinfo {author} {\bibfnamefont {X.-Y.}\ \bibnamefont {Chen}}, \bibinfo
  {author} {\bibfnamefont {M.}~\bibnamefont {Duda}}, \bibinfo {author}
  {\bibfnamefont {T.}~\bibnamefont {Karman}}, \bibinfo {author} {\bibfnamefont
  {I.}~\bibnamefont {Bloch}},\ and\ \bibinfo {author} {\bibfnamefont {X.-Y.}\
  \bibnamefont {Luo}},\ }\href {https://doi.org/10.1038/s41586-022-04900-0}
  {\bibfield  {journal} {\bibinfo  {journal} {Nature}\ }\textbf {\bibinfo
  {volume} {607}},\ \bibinfo {pages} {677} (\bibinfo {year}
  {2022})}\BibitemShut {NoStop}%
\bibitem [{\citenamefont {Bigagli}\ \emph {et~al.}(2023)\citenamefont
  {Bigagli}, \citenamefont {Warner}, \citenamefont {Yuan}, \citenamefont
  {Zhang}, \citenamefont {Stevenson}, \citenamefont {Karman},\ and\
  \citenamefont {Will}}]{Bigagli23_NatPhys}%
  \BibitemOpen
  \bibfield  {author} {\bibinfo {author} {\bibfnamefont {N.}~\bibnamefont
  {Bigagli}}, \bibinfo {author} {\bibfnamefont {C.}~\bibnamefont {Warner}},
  \bibinfo {author} {\bibfnamefont {W.}~\bibnamefont {Yuan}}, \bibinfo {author}
  {\bibfnamefont {S.}~\bibnamefont {Zhang}}, \bibinfo {author} {\bibfnamefont
  {I.}~\bibnamefont {Stevenson}}, \bibinfo {author} {\bibfnamefont
  {T.}~\bibnamefont {Karman}},\ and\ \bibinfo {author} {\bibfnamefont
  {S.}~\bibnamefont {Will}},\ }\bibfield  {journal} {\bibinfo  {journal}
  {Nature Physics}\ }\href {https://doi.org/10.1038/s41567-023-02200-6}
  {10.1038/s41567-023-02200-6} (\bibinfo {year} {2023})\BibitemShut {NoStop}%
\bibitem [{\citenamefont {Lin}\ \emph {et~al.}(2023)\citenamefont {Lin},
  \citenamefont {Chen}, \citenamefont {Jin}, \citenamefont {Shi}, \citenamefont
  {Deng}, \citenamefont {Zhang}, \citenamefont {Qu\'em\'ener}, \citenamefont
  {Shi}, \citenamefont {Yi},\ and\ \citenamefont {Wang}}]{Lin23_PRX}%
  \BibitemOpen
  \bibfield  {author} {\bibinfo {author} {\bibfnamefont {J.}~\bibnamefont
  {Lin}}, \bibinfo {author} {\bibfnamefont {G.}~\bibnamefont {Chen}}, \bibinfo
  {author} {\bibfnamefont {M.}~\bibnamefont {Jin}}, \bibinfo {author}
  {\bibfnamefont {Z.}~\bibnamefont {Shi}}, \bibinfo {author} {\bibfnamefont
  {F.}~\bibnamefont {Deng}}, \bibinfo {author} {\bibfnamefont {W.}~\bibnamefont
  {Zhang}}, \bibinfo {author} {\bibfnamefont {G.}~\bibnamefont {Qu\'em\'ener}},
  \bibinfo {author} {\bibfnamefont {T.}~\bibnamefont {Shi}}, \bibinfo {author}
  {\bibfnamefont {S.}~\bibnamefont {Yi}},\ and\ \bibinfo {author}
  {\bibfnamefont {D.}~\bibnamefont {Wang}},\ }\href
  {https://doi.org/10.1103/PhysRevX.13.031032} {\bibfield  {journal} {\bibinfo
  {journal} {Phys. Rev. X}\ }\textbf {\bibinfo {volume} {13}},\ \bibinfo
  {pages} {031032} (\bibinfo {year} {2023})}\BibitemShut {NoStop}%
\bibitem [{\citenamefont {Sadeghpour}\ \emph {et~al.}(2000)\citenamefont
  {Sadeghpour}, \citenamefont {Bohn}, \citenamefont {Cavagnero}, \citenamefont
  {Esry}, \citenamefont {Fabrikant}, \citenamefont {Macek},\ and\ \citenamefont
  {Rau}}]{Sadeghpour00_JPB}%
  \BibitemOpen
  \bibfield  {author} {\bibinfo {author} {\bibfnamefont {H.~R.}\ \bibnamefont
  {Sadeghpour}}, \bibinfo {author} {\bibfnamefont {J.~L.}\ \bibnamefont
  {Bohn}}, \bibinfo {author} {\bibfnamefont {M.~J.}\ \bibnamefont {Cavagnero}},
  \bibinfo {author} {\bibfnamefont {B.~D.}\ \bibnamefont {Esry}}, \bibinfo
  {author} {\bibfnamefont {I.~I.}\ \bibnamefont {Fabrikant}}, \bibinfo {author}
  {\bibfnamefont {J.~H.}\ \bibnamefont {Macek}},\ and\ \bibinfo {author}
  {\bibfnamefont {A.~R.~P.}\ \bibnamefont {Rau}},\ }\href
  {https://doi.org/10.1088/0953-4075/33/5/201} {\bibfield  {journal} {\bibinfo
  {journal} {Journal of Physics B: Atomic, Molecular and Optical Physics}\
  }\textbf {\bibinfo {volume} {33}},\ \bibinfo {pages} {R93} (\bibinfo {year}
  {2000})}\BibitemShut {NoStop}%
\bibitem [{\citenamefont {Aikawa}\ \emph {et~al.}(2014)\citenamefont {Aikawa},
  \citenamefont {Frisch}, \citenamefont {Mark}, \citenamefont {Baier},
  \citenamefont {Grimm}, \citenamefont {Bohn}, \citenamefont {Jin},
  \citenamefont {Bruun},\ and\ \citenamefont {Ferlaino}}]{Aikawa14_PRL}%
  \BibitemOpen
  \bibfield  {author} {\bibinfo {author} {\bibfnamefont {K.}~\bibnamefont
  {Aikawa}}, \bibinfo {author} {\bibfnamefont {A.}~\bibnamefont {Frisch}},
  \bibinfo {author} {\bibfnamefont {M.}~\bibnamefont {Mark}}, \bibinfo {author}
  {\bibfnamefont {S.}~\bibnamefont {Baier}}, \bibinfo {author} {\bibfnamefont
  {R.}~\bibnamefont {Grimm}}, \bibinfo {author} {\bibfnamefont {J.~L.}\
  \bibnamefont {Bohn}}, \bibinfo {author} {\bibfnamefont {D.~S.}\ \bibnamefont
  {Jin}}, \bibinfo {author} {\bibfnamefont {G.~M.}\ \bibnamefont {Bruun}},\
  and\ \bibinfo {author} {\bibfnamefont {F.}~\bibnamefont {Ferlaino}},\ }\href
  {https://doi.org/10.1103/PhysRevLett.113.263201} {\bibfield  {journal}
  {\bibinfo  {journal} {Phys. Rev. Lett.}\ }\textbf {\bibinfo {volume} {113}},\
  \bibinfo {pages} {263201} (\bibinfo {year} {2014})}\BibitemShut {NoStop}%
\bibitem [{\citenamefont {Tang}\ \emph {et~al.}(2016)\citenamefont {Tang},
  \citenamefont {Sykes}, \citenamefont {Burdick}, \citenamefont {DiSciacca},
  \citenamefont {Petrov},\ and\ \citenamefont {Lev}}]{Tang16_PRL}%
  \BibitemOpen
  \bibfield  {author} {\bibinfo {author} {\bibfnamefont {Y.}~\bibnamefont
  {Tang}}, \bibinfo {author} {\bibfnamefont {A.~G.}\ \bibnamefont {Sykes}},
  \bibinfo {author} {\bibfnamefont {N.~Q.}\ \bibnamefont {Burdick}}, \bibinfo
  {author} {\bibfnamefont {J.~M.}\ \bibnamefont {DiSciacca}}, \bibinfo {author}
  {\bibfnamefont {D.~S.}\ \bibnamefont {Petrov}},\ and\ \bibinfo {author}
  {\bibfnamefont {B.~L.}\ \bibnamefont {Lev}},\ }\href
  {https://doi.org/10.1103/PhysRevLett.117.155301} {\bibfield  {journal}
  {\bibinfo  {journal} {Phys. Rev. Lett.}\ }\textbf {\bibinfo {volume} {117}},\
  \bibinfo {pages} {155301} (\bibinfo {year} {2016})}\BibitemShut {NoStop}%
\bibitem [{\citenamefont {Tang}\ \emph {et~al.}(2015)\citenamefont {Tang},
  \citenamefont {Sykes}, \citenamefont {Burdick}, \citenamefont {Bohn},\ and\
  \citenamefont {Lev}}]{Tang16_PRA}%
  \BibitemOpen
  \bibfield  {author} {\bibinfo {author} {\bibfnamefont {Y.}~\bibnamefont
  {Tang}}, \bibinfo {author} {\bibfnamefont {A.}~\bibnamefont {Sykes}},
  \bibinfo {author} {\bibfnamefont {N.~Q.}\ \bibnamefont {Burdick}}, \bibinfo
  {author} {\bibfnamefont {J.~L.}\ \bibnamefont {Bohn}},\ and\ \bibinfo
  {author} {\bibfnamefont {B.~L.}\ \bibnamefont {Lev}},\ }\href
  {https://doi.org/10.1103/PhysRevA.92.022703} {\bibfield  {journal} {\bibinfo
  {journal} {Phys. Rev. A}\ }\textbf {\bibinfo {volume} {92}},\ \bibinfo
  {pages} {022703} (\bibinfo {year} {2015})}\BibitemShut {NoStop}%
\bibitem [{\citenamefont {Patscheider}\ \emph {et~al.}(2022)\citenamefont
  {Patscheider}, \citenamefont {Chomaz}, \citenamefont {Natale}, \citenamefont
  {Petter}, \citenamefont {Mark}, \citenamefont {Baier}, \citenamefont {Yang},
  \citenamefont {Wang}, \citenamefont {Bohn},\ and\ \citenamefont
  {Ferlaino}}]{Patscheider21_PRA}%
  \BibitemOpen
  \bibfield  {author} {\bibinfo {author} {\bibfnamefont {A.}~\bibnamefont
  {Patscheider}}, \bibinfo {author} {\bibfnamefont {L.}~\bibnamefont {Chomaz}},
  \bibinfo {author} {\bibfnamefont {G.}~\bibnamefont {Natale}}, \bibinfo
  {author} {\bibfnamefont {D.}~\bibnamefont {Petter}}, \bibinfo {author}
  {\bibfnamefont {M.~J.}\ \bibnamefont {Mark}}, \bibinfo {author}
  {\bibfnamefont {S.}~\bibnamefont {Baier}}, \bibinfo {author} {\bibfnamefont
  {B.}~\bibnamefont {Yang}}, \bibinfo {author} {\bibfnamefont {R.~R.~W.}\
  \bibnamefont {Wang}}, \bibinfo {author} {\bibfnamefont {J.~L.}\ \bibnamefont
  {Bohn}},\ and\ \bibinfo {author} {\bibfnamefont {F.}~\bibnamefont
  {Ferlaino}},\ }\href {https://doi.org/10.1103/PhysRevA.105.063307} {\bibfield
   {journal} {\bibinfo  {journal} {Phys. Rev. A}\ }\textbf {\bibinfo {volume}
  {105}},\ \bibinfo {pages} {063307} (\bibinfo {year} {2022})}\BibitemShut
  {NoStop}%
\bibitem [{\citenamefont {Bohn}\ \emph {et~al.}(2009)\citenamefont {Bohn},
  \citenamefont {Cavagnero},\ and\ \citenamefont {Ticknor}}]{Bohn09_NJP}%
  \BibitemOpen
  \bibfield  {author} {\bibinfo {author} {\bibfnamefont {J.~L.}\ \bibnamefont
  {Bohn}}, \bibinfo {author} {\bibfnamefont {M.}~\bibnamefont {Cavagnero}},\
  and\ \bibinfo {author} {\bibfnamefont {C.}~\bibnamefont {Ticknor}},\ }\href
  {https://doi.org/10.1088/1367-2630/11/5/055039} {\bibfield  {journal}
  {\bibinfo  {journal} {New Journal of Physics}\ }\textbf {\bibinfo {volume}
  {11}},\ \bibinfo {pages} {055039} (\bibinfo {year} {2009})}\BibitemShut
  {NoStop}%
\bibitem [{\citenamefont {Bohn}\ and\ \citenamefont {Jin}(2014)}]{Bohn14_PRA}%
  \BibitemOpen
  \bibfield  {author} {\bibinfo {author} {\bibfnamefont {J.~L.}\ \bibnamefont
  {Bohn}}\ and\ \bibinfo {author} {\bibfnamefont {D.~S.}\ \bibnamefont {Jin}},\
  }\href {https://doi.org/10.1103/PhysRevA.89.022702} {\bibfield  {journal}
  {\bibinfo  {journal} {Phys. Rev. A}\ }\textbf {\bibinfo {volume} {89}},\
  \bibinfo {pages} {022702} (\bibinfo {year} {2014})}\BibitemShut {NoStop}%
\bibitem [{\citenamefont {Sykes}\ and\ \citenamefont
  {Bohn}(2015)}]{Sykes15_PRA}%
  \BibitemOpen
  \bibfield  {author} {\bibinfo {author} {\bibfnamefont {A.~G.}\ \bibnamefont
  {Sykes}}\ and\ \bibinfo {author} {\bibfnamefont {J.~L.}\ \bibnamefont
  {Bohn}},\ }\href {https://doi.org/10.1103/PhysRevA.91.013625} {\bibfield
  {journal} {\bibinfo  {journal} {Phys. Rev. A}\ }\textbf {\bibinfo {volume}
  {91}},\ \bibinfo {pages} {013625} (\bibinfo {year} {2015})}\BibitemShut
  {NoStop}%
\bibitem [{\citenamefont {Wang}\ and\ \citenamefont {Bohn}(2021)}]{Wang21_PRA}%
  \BibitemOpen
  \bibfield  {author} {\bibinfo {author} {\bibfnamefont {R.~R.~W.}\
  \bibnamefont {Wang}}\ and\ \bibinfo {author} {\bibfnamefont {J.~L.}\
  \bibnamefont {Bohn}},\ }\href {https://doi.org/10.1103/PhysRevA.103.063320}
  {\bibfield  {journal} {\bibinfo  {journal} {Phys. Rev. A}\ }\textbf {\bibinfo
  {volume} {103}},\ \bibinfo {pages} {063320} (\bibinfo {year}
  {2021})}\BibitemShut {NoStop}%
\bibitem [{\citenamefont {Luiten}\ \emph {et~al.}(1996)\citenamefont {Luiten},
  \citenamefont {Reynolds},\ and\ \citenamefont {Walraven}}]{Luiten96_PRA}%
  \BibitemOpen
  \bibfield  {author} {\bibinfo {author} {\bibfnamefont {O.~J.}\ \bibnamefont
  {Luiten}}, \bibinfo {author} {\bibfnamefont {M.~W.}\ \bibnamefont
  {Reynolds}},\ and\ \bibinfo {author} {\bibfnamefont {J.~T.~M.}\ \bibnamefont
  {Walraven}},\ }\href {https://doi.org/10.1103/PhysRevA.53.381} {\bibfield
  {journal} {\bibinfo  {journal} {Phys. Rev. A}\ }\textbf {\bibinfo {volume}
  {53}},\ \bibinfo {pages} {381} (\bibinfo {year} {1996})}\BibitemShut
  {NoStop}%
\bibitem [{\citenamefont {Davis}\ \emph {et~al.}(1995)\citenamefont {Davis},
  \citenamefont {Mewes},\ and\ \citenamefont {Ketterle}}]{Davis95_APB}%
  \BibitemOpen
  \bibfield  {author} {\bibinfo {author} {\bibfnamefont {K.~B.}\ \bibnamefont
  {Davis}}, \bibinfo {author} {\bibfnamefont {M.-O.}\ \bibnamefont {Mewes}},\
  and\ \bibinfo {author} {\bibfnamefont {W.}~\bibnamefont {Ketterle}},\ }\href
  {https://doi.org/10.1007/BF01135857} {\bibfield  {journal} {\bibinfo
  {journal} {Applied Physics B}\ }\textbf {\bibinfo {volume} {60}},\ \bibinfo
  {pages} {155} (\bibinfo {year} {1995})}\BibitemShut {NoStop}%
\bibitem [{\citenamefont {Deng}\ \emph {et~al.}(2023)\citenamefont {Deng},
  \citenamefont {Chen}, \citenamefont {Luo}, \citenamefont {Zhang},
  \citenamefont {Yi},\ and\ \citenamefont {Shi}}]{Deng23_PRL}%
  \BibitemOpen
  \bibfield  {author} {\bibinfo {author} {\bibfnamefont {F.}~\bibnamefont
  {Deng}}, \bibinfo {author} {\bibfnamefont {X.-Y.}\ \bibnamefont {Chen}},
  \bibinfo {author} {\bibfnamefont {X.-Y.}\ \bibnamefont {Luo}}, \bibinfo
  {author} {\bibfnamefont {W.}~\bibnamefont {Zhang}}, \bibinfo {author}
  {\bibfnamefont {S.}~\bibnamefont {Yi}},\ and\ \bibinfo {author}
  {\bibfnamefont {T.}~\bibnamefont {Shi}},\ }\href
  {https://doi.org/10.1103/PhysRevLett.130.183001} {\bibfield  {journal}
  {\bibinfo  {journal} {Phys. Rev. Lett.}\ }\textbf {\bibinfo {volume} {130}},\
  \bibinfo {pages} {183001} (\bibinfo {year} {2023})}\BibitemShut {NoStop}%
\bibitem [{\citenamefont {Wang}\ and\ \citenamefont
  {Qu{\'e}m{\'e}ner}(2015)}]{Wang15_NJP}%
  \BibitemOpen
  \bibfield  {author} {\bibinfo {author} {\bibfnamefont {G.}~\bibnamefont
  {Wang}}\ and\ \bibinfo {author} {\bibfnamefont {G.}~\bibnamefont
  {Qu{\'e}m{\'e}ner}},\ }\href
  {https://iopscience.iop.org/article/10.1088/1367-2630/17/3/035015/meta}
  {\bibfield  {journal} {\bibinfo  {journal} {New Journal of Physics}\ }\textbf
  {\bibinfo {volume} {17}},\ \bibinfo {pages} {035015} (\bibinfo {year}
  {2015})}\BibitemShut {NoStop}%
\bibitem [{\citenamefont {Pitaevskii}\ \emph {et~al.}(2017)\citenamefont
  {Pitaevskii}, \citenamefont {Lifshitz},\ and\ \citenamefont
  {Sykes}}]{Pitaevskii17_ES}%
  \BibitemOpen
  \bibfield  {author} {\bibinfo {author} {\bibfnamefont {L.}~\bibnamefont
  {Pitaevskii}}, \bibinfo {author} {\bibfnamefont {E.}~\bibnamefont
  {Lifshitz}},\ and\ \bibinfo {author} {\bibfnamefont {J.}~\bibnamefont
  {Sykes}},\ }\href {https://books.google.com/books?id=uq-cDAAAQBAJ} {\emph
  {\bibinfo {title} {Course of Theoretical Physics: Physical Kinetics}}},\
  COURSE OF THEORETICAL PHYSICS\ (\bibinfo  {publisher} {Elsevier Science},\
  \bibinfo {year} {2017})\BibitemShut {NoStop}%
\bibitem [{\citenamefont {Bird}(1970)}]{Bird70_PF}%
  \BibitemOpen
  \bibfield  {author} {\bibinfo {author} {\bibfnamefont {G.~A.}\ \bibnamefont
  {Bird}},\ }\href {https://doi.org/10.1063/1.1692849} {\bibfield  {journal}
  {\bibinfo  {journal} {The Physics of Fluids}\ }\textbf {\bibinfo {volume}
  {13}},\ \bibinfo {pages} {2676} (\bibinfo {year} {1970})},\ \Eprint
  {https://arxiv.org/abs/https://aip.scitation.org/doi/pdf/10.1063/1.1692849}
  {https://aip.scitation.org/doi/pdf/10.1063/1.1692849} \BibitemShut {NoStop}%
\bibitem [{Note1()}]{Note1}%
  \BibitemOpen
  \bibinfo {note} {The differential cross section suffers innate convergence
  issues due to singularities in the scattering amplitude \cite {Bohn14_PRA}.
  Fortunately for us, forward scattering does not contribute toward
  cross-dimensional thermalization, of which we are concerned with in this
  Letter. We leave addressing these issues to a future manuscript.}\BibitemShut
  {Stop}%
\bibitem [{\citenamefont {Sacks}\ \emph {et~al.}(1989)\citenamefont {Sacks},
  \citenamefont {Schiller},\ and\ \citenamefont {Welch}}]{Sacks89_TF}%
  \BibitemOpen
  \bibfield  {author} {\bibinfo {author} {\bibfnamefont {J.}~\bibnamefont
  {Sacks}}, \bibinfo {author} {\bibfnamefont {S.~B.}\ \bibnamefont
  {Schiller}},\ and\ \bibinfo {author} {\bibfnamefont {W.~J.}\ \bibnamefont
  {Welch}},\ }\href {https://doi.org/10.1080/00401706.1989.10488474} {\bibfield
   {journal} {\bibinfo  {journal} {Technometrics}\ }\textbf {\bibinfo {volume}
  {31}},\ \bibinfo {pages} {41} (\bibinfo {year} {1989})}\BibitemShut {NoStop}%
\bibitem [{\citenamefont {Cui}\ and\ \citenamefont {Krems}(2016)}]{Cui16_JPB}%
  \BibitemOpen
  \bibfield  {author} {\bibinfo {author} {\bibfnamefont {J.}~\bibnamefont
  {Cui}}\ and\ \bibinfo {author} {\bibfnamefont {R.~V.}\ \bibnamefont
  {Krems}},\ }\href@noop {} {\bibfield  {journal} {\bibinfo  {journal} {Journal
  of Physics B: Atomic, Molecular and Optical Physics}\ }\textbf {\bibinfo
  {volume} {49}},\ \bibinfo {pages} {224001} (\bibinfo {year}
  {2016})}\BibitemShut {NoStop}%
\bibitem [{\citenamefont {Christianen}\ \emph {et~al.}(2019)\citenamefont
  {Christianen}, \citenamefont {Karman}, \citenamefont {Vargas-Hernández},
  \citenamefont {Groenenboom},\ and\ \citenamefont
  {Krems}}]{Christianen19_JCP}%
  \BibitemOpen
  \bibfield  {author} {\bibinfo {author} {\bibfnamefont {A.}~\bibnamefont
  {Christianen}}, \bibinfo {author} {\bibfnamefont {T.}~\bibnamefont {Karman}},
  \bibinfo {author} {\bibfnamefont {R.~A.}\ \bibnamefont {Vargas-Hernández}},
  \bibinfo {author} {\bibfnamefont {G.~C.}\ \bibnamefont {Groenenboom}},\ and\
  \bibinfo {author} {\bibfnamefont {R.~V.}\ \bibnamefont {Krems}},\ }\href
  {https://doi.org/10.1063/1.5082740} {\bibfield  {journal} {\bibinfo
  {journal} {The Journal of Chemical Physics}\ }\textbf {\bibinfo {volume}
  {150}},\ \bibinfo {pages} {064106} (\bibinfo {year} {2019})},\ \Eprint
  {https://arxiv.org/abs/https://pubs.aip.org/aip/jcp/article-pdf/doi/10.1063/1.5082740/13569165/064106\_1\_online.pdf}
  {https://pubs.aip.org/aip/jcp/article-pdf/doi/10.1063/1.5082740/13569165/064106\_1\_online.pdf}
  \BibitemShut {NoStop}%
\bibitem [{\citenamefont {Rasmussen}\ and\ \citenamefont
  {Williams}(2005)}]{Rasmussen05_MIT}%
  \BibitemOpen
  \bibfield  {author} {\bibinfo {author} {\bibfnamefont {C.~E.}\ \bibnamefont
  {Rasmussen}}\ and\ \bibinfo {author} {\bibfnamefont {C.~K.~I.}\ \bibnamefont
  {Williams}},\ }\href {https://doi.org/10.7551/mitpress/3206.001.0001} {\emph
  {\bibinfo {title} {{Gaussian Processes for Machine Learning}}}}\ (\bibinfo
  {publisher} {The MIT Press},\ \bibinfo {year} {2005})\BibitemShut {NoStop}%
\bibitem [{Note2()}]{Note2}%
  \BibitemOpen
  \bibinfo {note} {We utilize more points than is usually necessary for GP
  fitting in this study, so as to obtain more accurate results of subsequently
  computed quantities in this Letter. We also optimize the model's
  hyperparameters \cite {Yang20_NC} on top of just the kernel parameters. Even
  so, the Gaussian process model has issues faithfully reproducing the
  differential cross section around $\eta , \theta _s = 90^{\circ }$, known to
  have a discontinuity at threshold \cite {Bohn14_PRA}. Fortunately, this
  angular segment corresponds to forward scattering, which does not contribute
  to the cross-dimensional thermalization process of interest here. We ignore
  this issue until necessary for consideration in future works.}\BibitemShut
  {Stop}%
\bibitem [{\citenamefont {Monroe}\ \emph {et~al.}(1993)\citenamefont {Monroe},
  \citenamefont {Cornell}, \citenamefont {Sackett}, \citenamefont {Myatt},\
  and\ \citenamefont {Wieman}}]{Monroe93_PRL}%
  \BibitemOpen
  \bibfield  {author} {\bibinfo {author} {\bibfnamefont {C.~R.}\ \bibnamefont
  {Monroe}}, \bibinfo {author} {\bibfnamefont {E.~A.}\ \bibnamefont {Cornell}},
  \bibinfo {author} {\bibfnamefont {C.~A.}\ \bibnamefont {Sackett}}, \bibinfo
  {author} {\bibfnamefont {C.~J.}\ \bibnamefont {Myatt}},\ and\ \bibinfo
  {author} {\bibfnamefont {C.~E.}\ \bibnamefont {Wieman}},\ }\href
  {https://doi.org/10.1103/PhysRevLett.70.414} {\bibfield  {journal} {\bibinfo
  {journal} {Phys. Rev. Lett.}\ }\textbf {\bibinfo {volume} {70}},\ \bibinfo
  {pages} {414} (\bibinfo {year} {1993})}\BibitemShut {NoStop}%
\bibitem [{\citenamefont {Snoke}\ and\ \citenamefont
  {Wolfe}(1989)}]{Snoke89_PRB}%
  \BibitemOpen
  \bibfield  {author} {\bibinfo {author} {\bibfnamefont {D.~W.}\ \bibnamefont
  {Snoke}}\ and\ \bibinfo {author} {\bibfnamefont {J.~P.}\ \bibnamefont
  {Wolfe}},\ }\href {https://doi.org/10.1103/PhysRevB.39.4030} {\bibfield
  {journal} {\bibinfo  {journal} {Phys. Rev. B}\ }\textbf {\bibinfo {volume}
  {39}},\ \bibinfo {pages} {4030} (\bibinfo {year} {1989})}\BibitemShut
  {NoStop}%
\bibitem [{Note3()}]{Note3}%
  \BibitemOpen
  \bibinfo {note} {The Monte Carlo integration gives a $\lesssim 1\%$ error,
  which is mostly imperceptible in the log-linear plot.}\BibitemShut {Stop}%
\bibitem [{\citenamefont {DeMarco}\ \emph {et~al.}(1999)\citenamefont
  {DeMarco}, \citenamefont {Bohn}, \citenamefont {Burke}, \citenamefont
  {Holland},\ and\ \citenamefont {Jin}}]{DeMarco99_PRL}%
  \BibitemOpen
  \bibfield  {author} {\bibinfo {author} {\bibfnamefont {B.}~\bibnamefont
  {DeMarco}}, \bibinfo {author} {\bibfnamefont {J.~L.}\ \bibnamefont {Bohn}},
  \bibinfo {author} {\bibfnamefont {J.~P.}\ \bibnamefont {Burke}}, \bibinfo
  {author} {\bibfnamefont {M.}~\bibnamefont {Holland}},\ and\ \bibinfo {author}
  {\bibfnamefont {D.~S.}\ \bibnamefont {Jin}},\ }\href
  {https://doi.org/10.1103/PhysRevLett.82.4208} {\bibfield  {journal} {\bibinfo
   {journal} {Phys. Rev. Lett.}\ }\textbf {\bibinfo {volume} {82}},\ \bibinfo
  {pages} {4208} (\bibinfo {year} {1999})}\BibitemShut {NoStop}%
\bibitem [{\citenamefont {Surkov}\ \emph {et~al.}(1996)\citenamefont {Surkov},
  \citenamefont {Walraven},\ and\ \citenamefont {Shlyapnikov}}]{Surkov96_PRA}%
  \BibitemOpen
  \bibfield  {author} {\bibinfo {author} {\bibfnamefont {E.~L.}\ \bibnamefont
  {Surkov}}, \bibinfo {author} {\bibfnamefont {J.~T.~M.}\ \bibnamefont
  {Walraven}},\ and\ \bibinfo {author} {\bibfnamefont {G.~V.}\ \bibnamefont
  {Shlyapnikov}},\ }\href {https://doi.org/10.1103/PhysRevA.53.3403} {\bibfield
   {journal} {\bibinfo  {journal} {Phys. Rev. A}\ }\textbf {\bibinfo {volume}
  {53}},\ \bibinfo {pages} {3403} (\bibinfo {year} {1996})}\BibitemShut
  {NoStop}%
\bibitem [{\citenamefont {Wang}\ and\ \citenamefont
  {Bohn}(2022{\natexlab{a}})}]{Wang22_PRA}%
  \BibitemOpen
  \bibfield  {author} {\bibinfo {author} {\bibfnamefont {R.~R.~W.}\
  \bibnamefont {Wang}}\ and\ \bibinfo {author} {\bibfnamefont {J.~L.}\
  \bibnamefont {Bohn}},\ }\href {https://doi.org/10.1103/PhysRevA.106.023319}
  {\bibfield  {journal} {\bibinfo  {journal} {Phys. Rev. A}\ }\textbf {\bibinfo
  {volume} {106}},\ \bibinfo {pages} {023319} (\bibinfo {year}
  {2022}{\natexlab{a}})}\BibitemShut {NoStop}%
\bibitem [{\citenamefont {Wang}\ and\ \citenamefont
  {Bohn}(2022{\natexlab{b}})}]{Wang22_PRA2}%
  \BibitemOpen
  \bibfield  {author} {\bibinfo {author} {\bibfnamefont {R.~R.~W.}\
  \bibnamefont {Wang}}\ and\ \bibinfo {author} {\bibfnamefont {J.~L.}\
  \bibnamefont {Bohn}},\ }\href {https://doi.org/10.1103/PhysRevA.106.053307}
  {\bibfield  {journal} {\bibinfo  {journal} {Phys. Rev. A}\ }\textbf {\bibinfo
  {volume} {106}},\ \bibinfo {pages} {053307} (\bibinfo {year}
  {2022}{\natexlab{b}})}\BibitemShut {NoStop}%
\bibitem [{\citenamefont {Wang}\ and\ \citenamefont
  {Bohn}(2023{\natexlab{a}})}]{Wang23_PRA}%
  \BibitemOpen
  \bibfield  {author} {\bibinfo {author} {\bibfnamefont {R.~R.~W.}\
  \bibnamefont {Wang}}\ and\ \bibinfo {author} {\bibfnamefont {J.~L.}\
  \bibnamefont {Bohn}},\ }\href {https://doi.org/10.1103/PhysRevA.107.033321}
  {\bibfield  {journal} {\bibinfo  {journal} {Phys. Rev. A}\ }\textbf {\bibinfo
  {volume} {107}},\ \bibinfo {pages} {033321} (\bibinfo {year}
  {2023}{\natexlab{a}})}\BibitemShut {NoStop}%
\bibitem [{\citenamefont {Wang}\ and\ \citenamefont
  {Bohn}(2023{\natexlab{b}})}]{Wang23_PRA2}%
  \BibitemOpen
  \bibfield  {author} {\bibinfo {author} {\bibfnamefont {R.~R.~W.}\
  \bibnamefont {Wang}}\ and\ \bibinfo {author} {\bibfnamefont {J.~L.}\
  \bibnamefont {Bohn}},\ }\href {https://doi.org/10.1103/PhysRevA.108.013322}
  {\bibfield  {journal} {\bibinfo  {journal} {Phys. Rev. A}\ }\textbf {\bibinfo
  {volume} {108}},\ \bibinfo {pages} {013322} (\bibinfo {year}
  {2023}{\natexlab{b}})}\BibitemShut {NoStop}%
\bibitem [{\citenamefont {Lassabli\`ere}\ and\ \citenamefont
  {Qu\'em\'ener}(2022)}]{Lassabliere22_PRA}%
  \BibitemOpen
  \bibfield  {author} {\bibinfo {author} {\bibfnamefont {L.}~\bibnamefont
  {Lassabli\`ere}}\ and\ \bibinfo {author} {\bibfnamefont {G.}~\bibnamefont
  {Qu\'em\'ener}},\ }\href {https://doi.org/10.1103/PhysRevA.106.033311}
  {\bibfield  {journal} {\bibinfo  {journal} {Phys. Rev. A}\ }\textbf {\bibinfo
  {volume} {106}},\ \bibinfo {pages} {033311} (\bibinfo {year}
  {2022})}\BibitemShut {NoStop}%
\bibitem [{\citenamefont {Yang}\ and\ \citenamefont {Shami}(2020)}]{Yang20_NC}%
  \BibitemOpen
  \bibfield  {author} {\bibinfo {author} {\bibfnamefont {L.}~\bibnamefont
  {Yang}}\ and\ \bibinfo {author} {\bibfnamefont {A.}~\bibnamefont {Shami}},\
  }\href {https://doi.org/https://doi.org/10.1016/j.neucom.2020.07.061}
  {\bibfield  {journal} {\bibinfo  {journal} {Neurocomputing}\ }\textbf
  {\bibinfo {volume} {415}},\ \bibinfo {pages} {295} (\bibinfo {year}
  {2020})}\BibitemShut {NoStop}%
\bibitem [{\citenamefont {Dulieu}\ and\ \citenamefont
  {Osterwalder}(2017)}]{Dulieu17_RSC}%
  \BibitemOpen
  \bibfield  {author} {\bibinfo {author} {\bibfnamefont {O.}~\bibnamefont
  {Dulieu}}\ and\ \bibinfo {author} {\bibfnamefont {A.}~\bibnamefont
  {Osterwalder}},\ }\href {https://doi.org/10.1039/9781782626800} {\emph
  {\bibinfo {title} {{Cold Chemistry: Molecular Scattering and Reactivity Near
  Absolute Zero}}}}\ (\bibinfo  {publisher} {The Royal Society of Chemistry},\
  \bibinfo {year} {2017})\BibitemShut {NoStop}%
\bibitem [{\citenamefont {Idziaszek}\ \emph {et~al.}(2010)\citenamefont
  {Idziaszek}, \citenamefont {Qu\'em\'ener}, \citenamefont {Bohn},\ and\
  \citenamefont {Julienne}}]{Idziaszek10_PRA}%
  \BibitemOpen
  \bibfield  {author} {\bibinfo {author} {\bibfnamefont {Z.}~\bibnamefont
  {Idziaszek}}, \bibinfo {author} {\bibfnamefont {G.}~\bibnamefont
  {Qu\'em\'ener}}, \bibinfo {author} {\bibfnamefont {J.~L.}\ \bibnamefont
  {Bohn}},\ and\ \bibinfo {author} {\bibfnamefont {P.~S.}\ \bibnamefont
  {Julienne}},\ }\href {https://doi.org/10.1103/PhysRevA.82.020703} {\bibfield
  {journal} {\bibinfo  {journal} {Phys. Rev. A}\ }\textbf {\bibinfo {volume}
  {82}},\ \bibinfo {pages} {020703} (\bibinfo {year} {2010})}\BibitemShut
  {NoStop}%
\bibitem [{\citenamefont {Karman}(2023)}]{Karman23_JPCA}%
  \BibitemOpen
  \bibfield  {author} {\bibinfo {author} {\bibfnamefont {T.}~\bibnamefont
  {Karman}},\ }\href {https://doi.org/10.1021/acs.jpca.3c00797} {\bibfield
  {journal} {\bibinfo  {journal} {The Journal of Physical Chemistry A}\
  }\textbf {\bibinfo {volume} {127}},\ \bibinfo {pages} {2194} (\bibinfo {year}
  {2023})},\ \bibinfo {note} {pMID: 36825902},\ \Eprint
  {https://arxiv.org/abs/https://doi.org/10.1021/acs.jpca.3c00797}
  {https://doi.org/10.1021/acs.jpca.3c00797} \BibitemShut {NoStop}%
\bibitem [{\citenamefont {Johnson}(1973)}]{Johnson73_JCP}%
  \BibitemOpen
  \bibfield  {author} {\bibinfo {author} {\bibfnamefont {B.}~\bibnamefont
  {Johnson}},\ }\href
  {https://doi.org/https://doi.org/10.1016/0021-9991(73)90049-1} {\bibfield
  {journal} {\bibinfo  {journal} {Journal of Computational Physics}\ }\textbf
  {\bibinfo {volume} {13}},\ \bibinfo {pages} {445} (\bibinfo {year}
  {1973})}\BibitemShut {NoStop}%
\bibitem [{\citenamefont {Reif}(2009)}]{Reif09_Waveland}%
  \BibitemOpen
  \bibfield  {author} {\bibinfo {author} {\bibfnamefont {F.}~\bibnamefont
  {Reif}},\ }\href {https://www.waveland.com/browse.php?t=520} {\emph {\bibinfo
  {title} {Fundamentals of statistical and thermal physics}}}\ (\bibinfo
  {publisher} {Waveland Press},\ \bibinfo {address} {Long Grove, Illinois},\
  \bibinfo {year} {2009})\BibitemShut {NoStop}%
\end{thebibliography}%

\onecolumngrid
\clearpage

\begin{center}
{ \Large {\bf Supplemental material for: Prospects for thermalization of microwave-shielded ultracold molecules} } \vspace{10pt}

{ \large Reuben R. W. Wang and John L. Bohn } \vspace{10pt}

{ \normalsize {\it JILA, NIST, and Department of Physics, University of Colorado, Boulder, Colorado 80309, USA} }
\end{center}

\setcounter{page}{1}
\setcounter{equation}{0}

\section{ Scattering calculations of shielded molecules \label{app:scattering_calculations} }

For 2 polar molecules scattering of the effective potential $V_{\rm eff}(\boldsymbol{r})$ provided in the main text, scattering solutions can be obtained by first expanding the wavefunction in the basis
\begin{align}
    \psi( \boldsymbol{r} )
    =
    \sum_{\ell, m_{\ell}}
    \frac{ u_{E, \ell, m_{\ell}}(r) }{ r }
    Y_{\ell, m_{\ell}}(\theta, \phi),
\end{align}
where $Y_{\ell, m_{\ell}}(\theta, \phi)$ are spherical harmonics, and $u_{E, \ell, m_{\ell}}(r)$ are solutions to the radial time-independent Schr\"odinger equation:
\begin{align} \label{eq:radial_SE}
    \left(
    \frac{ d^2 }{ d r^2 }
    -
    \frac{ \ell ( \ell + 1 ) }{ r^2 }
    + 
    k^2
    \right)
    u_{E, \ell, m_{\ell}}(r) 
    =
    \frac{ 2 \mu }{ \hbar^2 } 
    \sum_{ \ell', m'_{\ell} }
    \bra{\ell, m_{\ell}} 
    V_{\rm eff}(\boldsymbol{{r}}) 
    \ket{\ell', m'_{\ell}}
    u_{E, \ell', m'_{\ell}}(r). 
\end{align}
Above, $k^2 = { 2 \mu E / \hbar^2 }$ is the collision wavenumber, and the explicit matrix elements $\bra{\ell, m_{\ell}} V_{\rm eff}(\boldsymbol{{r}}) \ket{\ell', m'_{\ell}}$, provided in below (\ref{app:matrix_elements}).

Numerical scattering solutions associated to Eq.~(\ref{eq:radial_SE}), require picking a consistent convention when referencing the associated scattering matrices. We present our adopted convention as follows. First defining the matrices
\begin{subequations}
\begin{align}
    D_{\ell, m_{\ell}}^{\ell', m'_{\ell}} 
    &=
    \delta_{\ell, \ell'}
    \delta_{m_{\ell}, m'_{\ell}}
    \frac{ d^2 }{ d r^2 }, \\
    W_{\ell, m_{\ell}}^{\ell', m'_{\ell}} 
    &=
    \delta_{\ell, \ell'}
    \delta_{m_{\ell}, m'_{\ell}}
    \left(
    k^2
    -
    \frac{ \ell ( \ell + 1 ) }{ r^2 }
    \right) 
    -
    \frac{ 2 \mu }{ \hbar^2 } 
    \bra{\ell, m_{\ell}} 
    V_{\rm eff}(\boldsymbol{{r}}) 
    \ket{\ell', m'_{\ell}},
\end{align}
\end{subequations}
and the fundamental set of radial wavefunction solutions $\boldsymbol{U}(r; E)$, Eq.~(\ref{eq:radial_SE}) can be recast as the compact system of equations:
\begin{align}
    \left[
    \boldsymbol{D} + \boldsymbol{W} 
    \right] 
    \boldsymbol{U}
    =
    \boldsymbol{0}. 
\end{align}
In principle, these equations can be solved numerically at a given collision energy $E$, by propagating the log derivative matrix 
\begin{align}
    \boldsymbol{Y}(r)
    =
    \boldsymbol{U}^{-1}(r)
    \frac{ \partial \boldsymbol{U}(r) }{ \partial r }
    =
    \frac{ \partial \log \boldsymbol{U}(r) }{ \partial r },
\end{align}
from $r = 0$ to $r \rightarrow \infty$. In practice, however, propagating to $\infty$ is not possible so we only do so up to $r = r_{\rm match}$, then match $\boldsymbol{Y}(r)$ to the asymptotic solutions where the distant colliders no longer interact. 
Moreover, we side step the issue of singularities at the origin by imposing a short-range boundary condition by starting the propagation at a minimum radius $r = r_{\min}$, then initializing the diagonal log-derivative matrix there as \cite{Wang15_NJP, Dulieu17_RSC} 
\begin{align}
    Y_{\ell, m_{\ell}}^{\ell, m_{\ell}} (r_{\min})
    =
    -i \sqrt{ W_{\ell, m_{\ell}}^{\ell, m_{\ell}}(r_{\min}) },
\end{align}
that assumes universal short-range loss. This boundary condition prevents dipolar scattering resonances \cite{Idziaszek10_PRA, Karman23_JPCA} which simplifies our current study.
Propagation is done with an adaptive radial step size version of Johnson's algorithm \cite{Johnson73_JCP}, utilizing 
\begin{subequations}
\begin{align}
    r_{\min}
    &= 
    100 a_0, \\
    r_{\rm match} 
    &=
    \sqrt{ \frac{ \hbar^2 L ( L + 1 ) }{ m E } }
    + 
    50 a_d,
\end{align}
\end{subequations}
where $a_0$ is the Bohr radius and $L$ is the largest value of $\ell$ utilized in the calculation. Typically, we utilize $L = 121$ or as many as is required for numerical convergence.  

The asymptotic solutions to Eq.~(\ref{eq:radial_SE}) arise by considering the domain where $r$ is much larger than the range of the potential, so that Eq.~(\ref{eq:radial_SE}) is well approximated as
\begin{align} \label{eq:asymptotic_radial_SE}
    \left(
    \frac{ d^2 }{ d r^2 }
    -
    \frac{ \ell ( \ell + 1 ) }{ r^2 }
    + 
    k^2 
    \right)
    u_{E, \ell, m_{\ell}}(r) 
    \approx 0.
\end{align}
This asymptotic radial equation is solved by the 2 independent solutions (up to arbitrary normalization):
\begin{subequations}
\begin{align}
    f_{E, \ell}(r) 
    &= 
    k r j_{\ell}(k r),
    \\
    g_{E, \ell}(r)
    &= 
    k r n_{\ell}(k r),
\end{align}
\end{subequations}
where $j_{\ell}(kr)$ and $n_{\ell}(kr)$ are the spherical Bessel and Neumann functions respectively. Then defining the matrices
\begin{subequations} \label{eq:asymptotic_matrices}
\begin{align}
    F_{\ell, m_{\ell}}^{\ell', m'_{\ell}}(r; E)
    &=
    \delta_{\ell, \ell'} \delta_{m_{\ell}, m'_{\ell}} f_{E, \ell}(r), \\
    G_{\ell, m_{\ell}}^{\ell', m'_{\ell}}(r; E) 
    &=
    \delta_{\ell, \ell'} \delta_{m_{\ell}, m'_{\ell}} g_{E, \ell}(r),
\end{align}
\end{subequations}
arbitrary solutions to Eq.~(\ref{eq:asymptotic_radial_SE}), and in fact Eq.~(\ref{eq:radial_SE}), can be written as
\begin{align}
    \boldsymbol{U}(r)
    =
    \boldsymbol{N}
    \left[
    \boldsymbol{F} (r)
    -
    \boldsymbol{K}
    \boldsymbol{G} (r)
    \right],
\end{align}
where $\boldsymbol{K}$ is the reactance matrix that is responsible for matching the numerical scattering solutions $\boldsymbol{U}$ to the asymptotic solutions in Eq.~(\ref{eq:asymptotic_matrices}) at $r = r_{\rm match}$. In particular, the off-diagonal elements of $\boldsymbol{K}$ provide information on the channel couplings that arise due to the interaction potential for a given incident collision channel. 
The matrix $\boldsymbol{N}$ is relevant only for normalization. 

The reactance matrix can be written in terms of the logarithmic derivative via
\begin{align} 
    \boldsymbol{K}
    &=
    \left.
    \frac{ \boldsymbol{F} (r)
    \boldsymbol{Y} (r)
    -
    \frac{ \partial }{ \partial r } \boldsymbol{F} (r) }{ \boldsymbol{G} (r)
    \boldsymbol{Y} (r)
    -
    \frac{ \partial }{ \partial r } \boldsymbol{G} (r) }
    \right|_{r = r_{\rm match}},
\end{align}
from which, we can then compute the other scattering matrices via the relations \cite{Bohn09_NJP}
\begin{subequations}
\begin{align} 
    \boldsymbol{S}
    &=
    \frac{ \boldsymbol{I}
    + 
    i
    \boldsymbol{K} }{ \boldsymbol{I}
    - 
    i
    \boldsymbol{K} }, 
    \\
    \boldsymbol{T}
    &=
    i ( \boldsymbol{S} - \boldsymbol{I} ). 
\end{align}
\end{subequations}
The scattering matrices above permit us to evaluate the scattering amplitude, noting that $m_{\ell}$ remains a good quantum number in these collision (App.~\ref{app:matrix_elements}), as
\begin{align}
    f_{\rm sc}( \boldsymbol{k}, \hat{\boldsymbol{k}}' ) 
    &= 
    -\frac{ 2 \pi }{ k }
    \sum_{ m_{\ell} }
    \sum_{\ell, \ell'}
    i^{\ell}
    Y_{\ell, m_{\ell}}^*(\hat{\boldsymbol{k}}) 
    T_{\ell, m_{\ell}}^{\ell', m_{\ell}}(k) 
    Y_{\ell', m_{\ell}}(\hat{\boldsymbol{k}}')
    i^{-\ell'},
\end{align}  
which gives the appropriately antisymmetrized elastic differential cross section \cite{Bohn14_PRA} via 
\begin{align} 
    {\cal D}_{\rm el}
    ( \boldsymbol{k}, \hat{\boldsymbol{k}}' ) 
    &=
    \frac{1}{2} 
    \abs{ f_{\rm sc}( \boldsymbol{k}, \hat{\boldsymbol{k}}' )
    -
    f_{\rm sc}( \boldsymbol{k}, -\hat{\boldsymbol{k}}' ) }^2, 
\end{align}
total cross section \cite{Dulieu17_RSC} 
\begin{align}
    \sigma( \boldsymbol{k} )
    &=
    \int d^2 \hat{\boldsymbol{k}}' 
    {\cal D}_{\rm el}
    ( \boldsymbol{k}, \hat{\boldsymbol{k}}' ) \nonumber\\ 
    &=
    \frac{ 4 \pi^2 }{ k^2 }
    \sum_{ m_{\ell} }
    \sum_{\tilde{\ell}, \ell, \ell'}
    i^{ \ell - \tilde{\ell} }
    Y_{\tilde{\ell}, m_{\ell}}(\hat{\boldsymbol{k}})
    Y_{\ell, m_{\ell}}^*(\hat{\boldsymbol{k}})
    \left[
    T_{\tilde{\ell}, m_{\ell}}^{\ell', m_{\ell}}(k) 
    \right]^*
    T_{\ell, m_{\ell}}^{\ell', m_{\ell}}(k),
\end{align} 
and integral total cross section
\begin{align}
    \overline{\sigma}
    &= 
    \frac{ 2 \pi }{ k^2 }
    \sum_{ m_{\ell} }
    \sum_{\ell, \ell'}
    \abs{ T_{\ell, m_{\ell}}^{\ell', m_{\ell}}(k) }^2. 
\end{align}

\subsection{ Matrix elements of effective potential \label{app:matrix_elements} }

To perform the scattering calculations on the effective single-channel microwave shielded potential energy surface, we are required to compute the $\bra{\ell, m} V_{\rm eff}(\boldsymbol{{r}}) \ket{\ell', m'_{\ell}}$ matrix elements. We list these elements explicitly in this section. 
The matrix elements for $V_{\rm dd}(\boldsymbol{r})$ are given as
\begin{align}
    \bra{\ell, m_{\ell}} V_{\rm dd}(\boldsymbol{r}) \ket{\ell', m'} 
    &=
    \frac{ d_{\rm eff}^2 }{ 4 \pi \epsilon_0 r^3 }
    \bra{\ell, m_{\ell}}
    \left[
    4 \sqrt{ \frac{ \pi }{ 5 } } Y_{2,0}(\theta, \phi) 
    \right] 
    \ket{\ell', m'_{\ell}} \nonumber\\
    &=
    \frac{ d_{\rm eff}^2 }{ 4 \pi \epsilon_0 r^3 }
    \Bigg[
    4 \sqrt{ \frac{ \pi }{ 5 } } 
    \int d\Omega 
    Y_{\ell, m_{\ell}}^*(\Omega) Y_{2,0}(\Omega) Y_{\ell', m'_{\ell}}(\Omega)  
    \Bigg] 
    \nonumber\\ 
    &=
    \frac{ d_{\rm eff}^2 }{ 4 \pi \epsilon_0 r^3 }
    2 ( -1 )^{m_{\ell}} \sqrt{ (2\ell + 1) (2\ell' + 1) }
    \begin{pmatrix}
        \ell & 2 & \ell' \\
        0 & 0 & 0
    \end{pmatrix}  
    \begin{pmatrix}
        \ell & 2 & \ell' \\
        -m_{\ell} & 0 & m'_{\ell}
    \end{pmatrix}, 
\end{align}
while the matrix elements for $V_{6}(\boldsymbol{r})$ are given as
\begin{align}
    \bra{\ell, m_{\ell}} V_{6}(\boldsymbol{r}) \ket{\ell', m'_{\ell}}
    &=
    \frac{ C_6 }{ r^6 }
    \bra{\ell, m_{\ell}} 
    \left(
    1
    +
    \cos^2\theta
    \right)
    \sin^2\theta 
    \ket{\ell', m'_{\ell}} \nonumber\\ 
    &=
    4 \sqrt{\pi } \frac{ C_6 }{ r^6 }
    \bra{\ell, m_{\ell}} 
    \bigg[
    \frac{ 2 }{ 5 }
    Y_{0,0}(\theta ,\phi ) 
    \frac{ 2 }{ 7 } \sqrt{\frac{1 }{5}} 
    Y_{2,0}(\theta ,\phi ) 
    -\frac{ 4 }{ 105 }
    Y_{4, 0}(\theta ,\phi ) 
    \bigg]
    \ket{\ell', m'_{\ell}} \nonumber\\
    &=
    \frac{ C_6 }{ r^6 } 
    2 ( -1 )^m
    \sqrt{ ( 2 \ell + 1 ) ( 2 \ell' + 1 ) } \nonumber\\
    &\quad\:\: \times 
    \bigg[
    \frac{ 2 }{ 5 } 
    \begin{pmatrix}
        \ell & 0 & \ell' \\
        0 & 0 & 0
    \end{pmatrix}
    \begin{pmatrix}
        \ell & 0 & \ell' \\
        -m_{\ell} & 0 & m'_{\ell}
    \end{pmatrix}
    \frac{ 2 }{ 7 }
    \begin{pmatrix}
        \ell & 2 & \ell' \\
        0 & 0 & 0
    \end{pmatrix}
    \begin{pmatrix}
        \ell & 2 & \ell' \\
        -m_{\ell} & 0 & m'_{\ell}
    \end{pmatrix} \nonumber\\
    &\quad\quad\quad 
    - \frac{ 4 }{ 35 }
    \begin{pmatrix}
        \ell & 4 & \ell' \\
        0 & 0 & 0
    \end{pmatrix}
    \begin{pmatrix}
        \ell & 4 & \ell' \\
        -m_{\ell} & 0 & m'_{\ell}
    \end{pmatrix} 
    \bigg].
\end{align}

\section{ Frame transformations for Gaussian process fitting \label{app:frame_transformations_for_GP} }

For efficient GP fitting of the elastic differential cross section, it is optimal to choose a coordinate frame whereby the symmetries are most conveniently handled.   
Naively, the differential cross section during a two-body collision involves 3 unit vectors: 1) the dipole alignment axis $\hat{\boldsymbol{{\cal E}}}$, 2) the incident relative momentum $\boldsymbol{k}$ and 3) the outgoing relative momentum $\boldsymbol{k}'$, therefore requiring 6 angular coordinates. This description is the case in a lab-frame (LF), where without loss of generality, we define it such that the dipole axis lies in its $x,z$-plane $\hat{\boldsymbol{{\cal E}}}_{\rm LF} = ( \sin\Theta, 0, \cos\Theta )^T$, and the other 2 vectors are given in terms of spherical coordinates as
\begin{subequations}
\begin{align}
    \hat{\boldsymbol{k}}  
    &=
    \begin{pmatrix}
        \sin\theta \cos\phi \\
        \sin\theta \sin\phi \\
        \cos\theta
    \end{pmatrix}, \\
    \hat{\boldsymbol{k}}'  
    &=
    \begin{pmatrix}
        \sin\theta' \cos\phi' \\
        \sin\theta' \sin\phi' \\
        \cos\theta'
    \end{pmatrix}. 
\end{align}
\end{subequations}

However, we can also define a dipole-frame (DF) which utilizes the dipole alignment direction as its $z$-axis, $\hat{\boldsymbol{{\cal E}}} = \hat{\boldsymbol{z}}_{\rm DF}$, while its $x$ axis is aligned to the plane in which both $\hat{\boldsymbol{{\cal E}}}$ and $\hat{\boldsymbol{k}}$ lie, so that
\begin{align}
    \hat{\boldsymbol{y}}_{\rm DF}
    &=
    \frac{ \hat{\boldsymbol{{\cal E}}} \cross \hat{\boldsymbol{k}} }{ | \hat{\boldsymbol{{\cal E}}} \cross \hat{\boldsymbol{k}} | }  
    \frac{ 1 }{ | \hat{\boldsymbol{{\cal E}}} \cross \hat{\boldsymbol{k}} | }
    \begin{pmatrix}
        -\cos\Theta \sin\theta \sin\phi \\
        \cos\Theta \sin\theta \cos\phi - \sin\Theta \cos\theta \\
        \sin\Theta \sin\theta \sin\phi
    \end{pmatrix}.
\end{align}
The remaining $\hat{\boldsymbol{x}}_{\rm DF}$ axis is then obtained with the cross product $\hat{\boldsymbol{x}}_{\rm DF} = \hat{\boldsymbol{y}}_{\rm DF} \cross \hat{\boldsymbol{z}}_{\rm DF}$. In the event where $\hat{\boldsymbol{{\cal E}}}$ and $\hat{\boldsymbol{k}}$ coincide, we simply choose 
\begin{align}
    \hat{\boldsymbol{x}}_{\rm DF}
    =
    \begin{pmatrix}
        \sin(\Theta + \pi/2) \\
        0 \\
        \cos(\Theta + \pi/2)
    \end{pmatrix},
\end{align}
and also $\hat{\boldsymbol{y}}_{\rm DF} = \hat{\boldsymbol{{\cal E}}} \cross \hat{\boldsymbol{x}}_{\rm DF}$. The differential cross section only cares about the relative angle between $\hat{\boldsymbol{k}}$ and $\hat{\boldsymbol{{\cal E}}}$, but not the vectors themselves. The dipole frame allows a convenient handling of this fact, we can simply write $\hat{\boldsymbol{k}} = ( \sin\eta, 0, \cos\eta )^T$ where $\eta = \cos^{-1} \hat{\boldsymbol{k}} \cdot \hat{\boldsymbol{{\cal E}}}$. So to obtain the post-collision relative momentum in the dipole-frame, we can construct the required rotation matrix $\boldsymbol{R}( {\rm DF} \leftarrow {\rm LF} )$, by the method of direction cosines 
\begin{align}
    \boldsymbol{R}( {\rm DF} \leftarrow {\rm LF} )
    &=
    \begin{pmatrix}
        \hat{\boldsymbol{x}}_{\rm DF} \cdot \hat{\boldsymbol{x}}_{\rm LF} & 
        \hat{\boldsymbol{x}}_{\rm DF} \cdot \hat{\boldsymbol{y}}_{\rm LF} & 
        \hat{\boldsymbol{x}}_{\rm DF} \cdot \hat{\boldsymbol{z}}_{\rm LF} \\
        \hat{\boldsymbol{y}}_{\rm DF} \cdot \hat{\boldsymbol{x}}_{\rm LF} & 
        \hat{\boldsymbol{y}}_{\rm DF} \cdot \hat{\boldsymbol{y}}_{\rm LF} & 
        \hat{\boldsymbol{y}}_{\rm DF} \cdot \hat{\boldsymbol{z}}_{\rm LF} \\
        \hat{\boldsymbol{z}}_{\rm DF} \cdot \hat{\boldsymbol{x}}_{\rm LF} & 
        \hat{\boldsymbol{z}}_{\rm DF} \cdot \hat{\boldsymbol{y}}_{\rm LF} & 
        \hat{\boldsymbol{z}}_{\rm DF} \cdot \hat{\boldsymbol{z}}_{\rm LF}
    \end{pmatrix}. 
\end{align} 
The outbound relative collision vector is then given in dipole-frame as $\hat{\boldsymbol{k}}'( {\rm DF} ) = \boldsymbol{R}( {\rm DF} \leftarrow {\rm LF} )\hat{\boldsymbol{k}}'( {\rm LF} )$.
We then denote inclination and azimuthal scattering angles in the dipole frame as $\theta_s$ and $\phi_s$ respectively. 

Furthermore, the dipole frame as defined makes the differential cross section symmetric about the $x,z$-plane, only requiring us to specify $\phi_s$ within the range $[0, \pi]$. The entire differential cross section can then be obtained by specifying its value in the appropriate energy interval, and for $\eta, \theta_s, \phi_s \in [0, \pi]$.

\section{ The Eikonal approximation \label{app:Eikonal_approximation} }

At collision energies much larger than $E_{\rm dd}$, the scattering becomes semiclassical with the total cross section well approximated by the Eikonal approximation \cite{Bohn09_NJP}. Within this approximation, the scattering amplitude, considering on the $1/r^3$ long-range tail of $V_{\rm eff}$, is given by
\begin{align}
    f_{\rm sc}^{\rm Ei}( \boldsymbol{k}, \hat{\boldsymbol{k}}' )
    &=
    \frac{ a_d \tilde{k} }{ 2 \pi i }
    \int_{0}^{2\pi} d\phi \int_{0}^{\infty} d \tilde{b} 
    \: \tilde{b} e^{ i \tilde{q} \tilde{b} \cos\phi } 
    \left[ 
    \exp\left( -\frac{ 1 }{ \tilde{k} } \int_{-\infty}^{\infty} \tilde{V}_{\rm eff}(\tilde{\boldsymbol{r}}') d\tilde{z}' \right) 
    -
    1 \right] \nonumber\\
    &=
    \frac{ a_d \tilde{k} }{ 2 \pi i }
    \int \tilde{b}  d \tilde{b} d\phi
    e^{ i \tilde{q} \tilde{b} \cos\phi } 
    \left[ 
    \exp\left( -\frac{ 2 i }{ \tilde{k} \tilde{b}^2 } \sin^2\alpha \cos( 2 \phi - 2 \beta ) \right) 
    -
    1 \right], \nonumber
\end{align}
where $\alpha = \cos^{-1}( \hat{\boldsymbol{k}} \cdot \hat{\boldsymbol{{\cal E}}} )$, $\beta = \tan^{-1}( \hat{\boldsymbol{y}} \cdot \hat{\boldsymbol{{\cal E}}} / \hat{\boldsymbol{x}} \cdot \hat{\boldsymbol{{\cal E}}} )$, $\boldsymbol{q} = \boldsymbol{k} - \boldsymbol{k}'$ is the momentum transfer vector, $b$ is the impact parameter and tildes denote adimensional quantities normalized by the relevant dipole units (see the main text).  

From the scattering amplitude, we can compute the total cross section using the optical theorem
\begin{align}
    \sigma^{\rm Ei}_{\rm total}( \boldsymbol{k} )
    &=
    \frac{ 4 \pi }{ k }
    {\rm Im}
    \left\{
    f_{\rm sc}^{\rm Ei}( \tilde{q} = 0 )
    \right\},
\end{align}
which requires evaluation of the scattering amplitude at forward scattering $\tilde{q} = 0$:
\begin{align}
    f_{\rm sc}^{\rm Ei}( \boldsymbol{k}, \hat{\boldsymbol{k}}' )
    &=
    \frac{ a_d }{ 2 \pi i \tilde{k} }
    \int \tilde{\ell} d \tilde{\ell} d\phi 
    \left[ 
    e^{ -\frac{ 2 i \tilde{k} }{ \tilde{\ell}^2 } \sin^2\alpha \cos( 2 \phi - 2 \beta ) }  
    -
    1 \right] \nonumber\\
    &=
    \frac{ a_d }{ 2 \pi i \tilde{k} }
    \int \tilde{\ell} d \tilde{\ell} d\phi 
    \Bigg[ 
    \sum_{m=0}^{\infty}
    \epsilon_m 
    (-i)^m
    J_m \left( \frac{ 2 \tilde{k} }{ \tilde{\ell}^2 } \sin^2\alpha \right)
    \cos[ 2 m ( \phi - \beta ) ]
    -
    1 
    \Bigg] \nonumber\\
    &=
    \frac{ a_d }{ i \tilde{k} }
    \int \tilde{\ell} d \tilde{\ell} 
    \Bigg[ 
    J_0\left( \frac{ 2 \tilde{k} }{ \tilde{\ell}^2 } \sin^2\alpha \right) 
    -
    1 
    \Bigg] \nonumber\\
    &=
    i a_d \sin^2\alpha, 
\end{align}
where $\epsilon_m = 1$ if $m = 0$, and $\epsilon_m = 2$ if $m > 0$, and we utilized the substitution $\ell = k b$. Then plugging the forward scattering amplitude into the optical theorem gives
\begin{align}
    \sigma^{\rm Ei}_{\rm total}( \boldsymbol{k} )
    =
    \frac{ 4 \pi a_d }{ k }
    \left[
    1 - \big( \hat{\boldsymbol{k}} \cdot \hat{\boldsymbol{{\cal E}}} \big)^2
    \right],
\end{align}
which averaged over incident directions, gives
\begin{align}
    \overline{\sigma}^{\rm Ei}_{\rm total}
    =
    \frac{ 8 \pi a_d }{ 3 k },
\end{align}
identical to the formula obtained for point-dipole scattering in Ref.~\cite{Bohn09_NJP}. The result above is applicable to distinguishable dipoles, and so may not be quantitatively accurate in describing our study of scattering between identical fermions. Nevertheless, it serves to provide a useful visual aid to the expectation energy scaling, and seems to actually give rather favorable quantitative agreement.

\section{ Deriving the collisional efficiency toward thermalization \label{app:CET} }

Obtaining the form of ${\cal N}_{i j}$ in the main text requires formulation of the Enskog equations. To do so, we define the phase space averaged quantity $\langle \chi_i \rangle = k_B ( {\cal T}_i - T_{\rm eq} )$, which quantifies the system's deviation from its equilibration temperature, $T_{\rm eq}$. Then multiplying the Boltzmann equation \cite{Reif09_Waveland} by $\chi_i$ and integrating it over phase space variables \cite{Wang21_PRA}, we derived Enskog equations that govern the relaxation of $\langle \chi_j \rangle$:
\begin{subequations}
\begin{align} 
    \frac{ d \langle \chi_i \rangle }{ dt } 
    &= 
    {\cal C}[ \chi_i ], \label{eq:Enskog_equation} 
    \\
    {\cal C}[ \chi_i ] 
    &=
    \frac{ \langle n \rangle }{ 2 }
    \int \frac{ d^3 \boldsymbol{p}_r }{ m }
    p_r
    c_r(\boldsymbol{p}_r, t) 
    \int d^2\Omega' 
    {\cal D}_{\rm el}
    \Delta\chi_i, \label{eq:Enskog_collision_integral}
\end{align}
\end{subequations} 
where $\langle n \rangle$ is the average number density, $c_r(\boldsymbol{p}_r, t)$ is the distribution of relative momentum $\boldsymbol{p}_r$, and $\Delta\chi \equiv \chi^{\prime} + \chi_1^{\prime} - \chi - \chi_1$ denotes the amount by which $\chi$ changes during a collision event. 

Taken only perturbatively from equilibrium along axis $i$, Eq.~\ref{eq:Enskog_equation} is approximated by the decay law ${\cal C}[ \chi_j ] \approx - \gamma_{i j} \langle \chi_j \rangle$, which results in the short-time relation
\begin{align} \label{eq:retherm_rate_theory}
    \gamma_{i j}
    =
    - \left. 
    \frac{ 1 }{ \left( \mathcal{T}_j(t) - T_{\text{eq}} \right) } \frac{ d \mathcal{T}_j(t) }{ d t }
    \right|_{t = 0}, 
\end{align}
identifying $\gamma_{i j}$ as the thermalization rate.  
Now considering the collision integral
\begin{align}
    {\cal C}[ \chi_j ] 
    &= 
    \frac{ \langle n \rangle }{ 2 } 
    \int \frac{ d^3 \boldsymbol{p}_r }{ m }
    p_r
    c_r(\boldsymbol{p}_r, t) 
    \int d^2\Omega'  
    {\cal D}_{\rm el}
    \Delta\chi_j, \nonumber\\
    &=
    \frac{ k_B \langle n \rangle }{ 2 }
    \int \frac{ d^3 \boldsymbol{p}_r }{ m }
    p_r
    c_r(\boldsymbol{p}_r, t) 
    \int d^2\Omega'  
    {\cal D}_{\rm el}
    \Delta {\cal T}_j, 
\end{align}
we move to center of mass and relative momentum coordinates, $\boldsymbol{P} = (\boldsymbol{p} + \boldsymbol{p}_1) / 2$ and $\boldsymbol{p}_r = \boldsymbol{p} - \boldsymbol{p}_1$ respectively, so that the change in $\chi_i$ is given as
\begin{align} 
    \Delta \chi_i
    &= 
    \frac{ \Delta p_i^2  }{ 2 m } 
    \nonumber\\
    &= 
    \frac{ p'^2_{j} + p'^2_{1, j} - p_{j}^2 - p_{1, j}^2 }{ 2 m } \nonumber\\
    &= 
    \frac{ p'^2_{r, j} + P'^2_{j} - p^2_{r, j} - P^2_{j} }{ 4 m } \nonumber\\
    &= 
    \frac{ p'^2_{r, i} - p^2_{r, i} }{ 4 m }.
\end{align} 
As for the relative momentum distribution, we Taylor expand it at $t = 0$ with respect to $\delta_i / k_B T_0$ to get
\begin{subequations}
\begin{align}
    c_r( \boldsymbol{p}_r, 0 )
    &=
    \prod_i 
    \left( \frac{ 1 }{ 4 \pi m k_B T_i } \right)^{1/2}
    \exp\left(
    -\frac{ p_{r,i}^2 }{ 4 m k_B T_i }
    \right) \nonumber\\
    &\approx 
    c_r^{(0)}(\boldsymbol{p}_r)
    \left[
    1
    +
    \left( \frac{ p_{r, i}^2 }{ 4 m k_B T_0 } - \frac{ 1 }{ 2 } \right)
    \frac{ \delta_i }{ k_B T_0 }
    \right], \\
    c_r^{(0)}(p_r)
    &=
    \frac{ 1 }{ ( 4 \pi m k_B T_0 )^{3/2} }
    \exp\left(
    -\frac{ p_{r}^2 }{ 4 m k_B T_0 }
    \right).
\end{align}
\end{subequations}

The expressions above render the collision integral 
\begin{align}
    {\cal C}[ \chi_j ]  
    &\approx
    \frac{ \langle n \rangle }{ 2 } 
    \int \frac{ d^3 \boldsymbol{p}_r }{ m }
    p_r
    c_r^{(0)}(\boldsymbol{p}_r)
    \left[
    1
    +
    \left( \frac{ p_{r, i}^2 }{ 4 m k_B T_0 } - \frac{ 1 }{ 2 } \right)
    \frac{ \delta_i }{ k_B T_0 }
    \right]
    \int d^2\Omega' 
    {\cal D}_{\rm el} 
    \left(
    \frac{ p'^2_{r, j} - p^2_{r, j} }{ 4 m }
    \right) \nonumber\\
    &=
    \frac{ \delta_i }{ 16 ( m k_B T_0 )^2 }
    \frac{ \langle n \rangle }{ 2 } 
    \int \frac{ d^3 \boldsymbol{p}_r }{ m }
    c_r^{(0)}(\boldsymbol{p}_r)
    p_r
    \int d^2\Omega' 
    {\cal D}'_{\rm el} 
    p_{r, i}^2
    \left(
    p'^2_{r, j} - p^2_{r, j}
    \right), 
\end{align}
which upon utilizing the time-reversal symmetry of elastic collisions
\begin{align}
    {\cal C}[ \chi_j ] 
    &\approx 
    \frac{ \delta_i }{ 16 ( m k_B T_0 )^2 }
    \frac{ \langle n \rangle }{ 2 } 
    \int p_r^2 d p_r
    c_r^{(0)}(p_r)
    \frac{ p_r }{ m }
    \int d^2\Omega d^2\Omega' 
    {\cal D}'_{\rm el} 
    p_{r, i}^2
    \left(
    p'^2_{r, j} - p^2_{r, j}
    \right) \nonumber\\
    &=
    \frac{ \delta_i }{ 16 ( m k_B T_0 )^2 }
    \frac{ \langle n \rangle }{ 2 } 
    \int p_r^2 d p_r
    c_r^{(0)}(p_r)
    \frac{ p_r }{ m }
    \int d^2\Omega' d^2\Omega 
    {\cal D}'_{\rm el} 
    p_{r, i}'^2
    \left(
    p^2_{r, j} - p'^2_{r, j}
    \right),
\end{align}
the expression above can also be written in a form that is explicit in the symmetry under exchange of indices $i$ and $j$:
\begin{align}
    {\cal C}[ \chi_j ]
    &=
    -\frac{ \delta_i }{ 32 ( m k_B T_0 )^2 }
    \frac{ \langle n \rangle }{ 2 } 
    \int \frac{ d^3 \boldsymbol{p}_r }{ m }
    c_r^{(0)}(p_r)
    p_r
    \int d^2\Omega' 
    {\cal D}'_{\rm el}
    \left(
    p'^2_{r, i} - p^2_{r, i}
    \right)
    \left(
    p'^2_{r, j} - p^2_{r, j}
    \right).
\end{align}
We have used the suggestive notation ${\cal D}'_{\rm el} = {\cal D}_{\rm el}(p_r, \Omega')$.
Plugging ${\cal C}[ \chi_j ]$ as written into Eq.~(\ref{eq:retherm_rate_theory}) and taking $\mathcal{T}_j(t) - T_{\rm eq} = \epsilon_j / k_B$, we obtain 
\begin{align} 
    \gamma_{i j}
    &=
    - \frac{ {\cal C}[ \mathcal{T}_j ] }{ \left( \mathcal{T}_j(t) - T_{\rm eq} \right) } \nonumber\\
    &=
    - \frac{ k_B }{ \epsilon_j }
    {\cal C}[ \mathcal{T}_j ] \nonumber\\
    &=
    \frac{ \delta_i }{ \epsilon_j } 
    \frac{ \langle n \rangle }{ 512 } 
    \int 
    \frac{ p_r^2 d p_r }{ ( \pi m k_B T_0 )^{3/2} }
    \exp\left(
    -\frac{ p_{r}^2 }{ 4 m k_B T_0 }
    \right)
    \frac{ p_r }{ m }
    \int 
    d^2\Omega d^2\Omega' 
    {\cal D}'_{\rm el}
    \left(
    \frac{ p'^2_{r, i} - p^2_{r, i} }{ m k_B T_0 }
    \right)
    \left(
    \frac{ p'^2_{r, j} - p^2_{r, j} }{ m k_B T_0 }
    \right).
\end{align} 
Finally, taking the limit of $\delta_i / (k_B T_0) \rightarrow 0$,   
we obtain $\varepsilon_{i j}$ in the main text, 
having defined $\alpha_{i j} = { \delta_i / \epsilon_j }$ and using the equipartition theorem ($T_{\rm eq} = T_0 + \delta_i/3 k_B$) to get
\begin{align}
    \frac{ \delta_i }{ \epsilon_j }
    =
    \begin{cases}
        { 3 / 2 }, & \quad i = j, \\
        -3, & \quad i \neq j.
    \end{cases}
\end{align}


\end{document}